\documentclass[journal]{IEEEtran}
\usepackage{array}
\usepackage[caption=false,font=normalsize,labelfont=sf,textfont=sf]{subfig}
\usepackage{textcomp}
\usepackage{stfloats}
\usepackage{verbatim}
\usepackage{graphicx}
\usepackage{cite}
\usepackage[flushleft]{threeparttable}
\usepackage{color}
\usepackage{booktabs}
\usepackage{multirow}
\usepackage{amsthm}
\usepackage{amsfonts}
\usepackage[ruled,norelsize]{algorithm2e}
\usepackage{hyperref}
\hypersetup{
	colorlinks=true,
	linkcolor=black,
	citecolor=black,
	urlcolor=black,
}

\newtheorem{definition}{Definition}
\newtheorem{theorem}{Theorem}
\renewcommand{\IEEEQED}{\IEEEQEDopen}

\begin{document}

\title{EfficientTDNN: Efficient Architecture Search for Speaker Recognition}

\author{Rui Wang,
        Zhihua Wei,
        Haoran Duan,~\IEEEmembership{Student Member, IEEE},
        Shouling Ji,~\IEEEmembership{Member, IEEE},\\
        Yang Long,
        Zhen Hong
\thanks{\emph{(Corresponding author: Zhihua Wei)}}
\thanks{Rui Wang, Zhihua Wei are with the Department of Computer Science and Technology, Tongji University, Shanghai 201804, China (e-mail: \{rwang,zhihua\_wei\}@tongji.edu.cn)}
\thanks{Haoran Duan is with the Department of Computer Science, Durham University, Durham, DH1 3LE, United Kingdom. (email: haoran.duan@ieee.org)}
\thanks{Shouling Ji is with the NESA Lab, College of Computer Science and Technology, Zhejiang University, Hangzhou 310027, China (e-mail: sji@zju.edu.cn)}
\thanks{Yang Long is with the Department of Computer Science, Durham University, Durham, DH1 3LE, United Kingdom. (e-mail: yang.long@durham.ac.uk)}
\thanks{Zhen Hong is with the Institute of Cyberspace Security, and the College of Information Engineering, Zhejiang University of Technology, Hangzhou 310023, China. (e-mail: zhong1983@zjut.edu.cn)}
\thanks{Manuscript received 24 November, 2021; revised 04 April, 2022.}}

\markboth{Manuscript Draft of IEEE/ACM Transactions on Audio, Speech, and Language Processing,~Vol. X, No.~X,X,X}%
{Wang \MakeLowercase{\textit{et al.}}: EfficientTDNN: Efficient Architecture Search for Speaker Recognition}

\IEEEpubid{0000--0000/00\$00.00~\copyright~2022 IEEE}



\maketitle

\begin{abstract}
Convolutional neural networks (CNNs), such as the time-delay neural network (TDNN), have shown their remarkable capability in learning speaker embedding. However, they meanwhile bring a huge computational cost in storage size, processing, and memory. Discovering the specialized CNN that meets a specific constraint requires a substantial effort of human experts. Compared with hand-designed approaches, neural architecture search (NAS) appears as a practical technique in automating the manual architecture design process and has attracted increasing interest in spoken language processing tasks such as speaker recognition.
In this paper, we propose EfficientTDNN, an efficient architecture search framework consisting of a TDNN-based supernet and a TDNN-NAS algorithm. The proposed supernet introduces temporal convolution of different ranges of the receptive field and feature aggregation of various resolutions from different layers to TDNN. On top of it, the TDNN-NAS algorithm quickly searches for the desired TDNN architecture via weight-sharing subnets, which surprisingly reduces computation while handling the vast number of devices with various resources requirements.
Experimental results on the VoxCeleb dataset show the proposed EfficientTDNN enables approximate $10^{13}$ architectures concerning depth, kernel, and width. Considering different computation constraints, it achieves a 2.20\% equal error rate (EER) with 204M multiply-accumulate operations (MACs), 1.41\% EER with 571M MACs as well as 0.94\% EER with 1.45G MACs. Comprehensive investigations suggest that the trained supernet generalizes subnets not sampled during training and obtains a favorable trade-off between accuracy and efficiency.
\end{abstract}


%


\IEEEpeerreviewmaketitle

\section{Introduction}\label{introduction}
Speaker recognition aims to identify the voice of the specific targets. 
For speaker recognition, convolutional neural networks (CNNs), including residual neural network (ResNet) \cite{Garcia-Romero2020} and time-delay neural networks (TDNN) \cite{Desplanques2020}, have shown their remarkable capability on learning speaker embedding and achieved impressive results. The performance of CNNs always comes with a considerable computational cost in terms of storage size, processing, and memory. 
It is essential to explore effective methods in designing neural architectures given the limited computing capacity such as mobile devices.

Discovering the specialized convolutional neural network that meets a certain constraint requires a substantial effort of human experts. Compared with hand-designed approaches, neural architecture search (NAS) \cite{Zoph2017} has made great achievements in the design of deep neural networks to automate the manual process of architecture design. Specifically, weight sharing NAS \cite{pmlr-v80-bender18a} can train a single large network (i.e., supernet) capable of emulating any architecture (i.e., subnet) in the search space. Recently, there have been increasing interests in exploring NAS for spoken language processing tasks, such as speech recognition \cite{Chen2020,Zheng2021,Mehrotra2021}, speech synthesis \cite{NEURIPS2020_77305c2f,Luo2021}, and speaker recognition \cite{Ding2020,Qu2020}.

\IEEEpubidadjcol

\begin{figure}[!t]
  \centering
  \includegraphics[width=8cm]{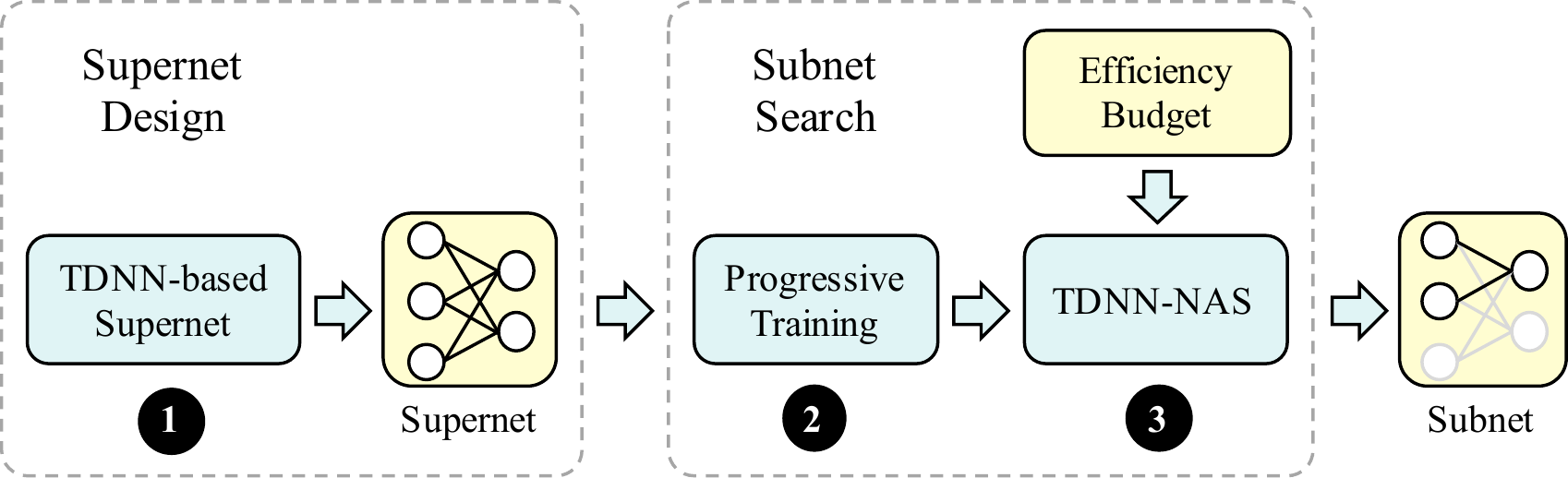}
  \caption{The proposed EfficientTDNN.}
  \label{fig:EfficientTDNN}
\end{figure}

However, applying NAS to speaker recognition to find the CNN that fits the specific constraint remains not well explored. There are two challenges: (1) the existing supernets are deficient in some of the inductive biases inherent to speaker networks, such as temporal convolution and feature aggregation; (2) the practicality of searching a subnet on large-scale datasets is discounted since retraining the standalone models may cause prohibitive computation. 

Inspired by once-for-all (OFA) \cite{Cai2020}, we attempt to propose a novel NAS framework to achieve efficient architecture search for speaker recognition.
In this paper, we propose an efficient architecture search framework, EfficientTDNN. As shown in Figure \ref{fig:EfficientTDNN}, the proposed EfficientTDNN consists of a TDNN-based supernet capable of emulating various TDNN architectures and a TDNN-NAS algorithm that quickly finds the desired TDNN.
Specifically, the proposed TDNN-based supernet introduces temporal convolution of different ranges of the receptive field and feature aggregation of various resolutions from different layers to TDNN, which improves the diversity of locality to boost the capabilities of discriminating speakers. The supernet is nested TDNN subnets that mutually share weights, which helps transfer knowledge between subnets of different sizes. 
On the top of the proposed supernet, the TDNN-NAS is proposed to search for the desired architecture without retraining, which significantly reduces computation compared with \cite{Ding2020} and \cite{Qu2020} while handling the vast number of devices with various resources requirements.
In the proposed EfficientTDNN, different-grained channel numbers in the search space allow exploring efficient architectures; progressive architectural shrinking is introduced for training with dynamic network layers, kernel sizes, and channel numbers.
Experiments are conducted on the VoxCeleb dataset show that the proposed EfficientTDNN enables approximate $10^{13}$ architectures concerning depth, kernel, and width and attains superior results under different computation constraints in that it achieves a 2.20\% equal error rate (EER) with 204M multiply-accumulate operations (MACs), 1.41\% EER with 571M MACs, and 0.94\% EER with 1.45G MACs. The models and code are released\footnote{\url{https://github.com/mechanicalsea/sugar}} to facilitate future research.

The contributions of this paper are summarized as follows.

\begin{itemize}
\item We propose EfficientTDNN as a NAS-powered alternative to automatically design efficient architectures given the specific computing capacity for speaker recognition. It consists of supernet design and subnets search.
\item We present a novel TDNN-based supernet that aggregates shallow-resolution and deep-resolution features with different ranges of the receptive field. It enables to sample subnets with different sizes in terms of depth, kernel, and width, which is essential to search for subnets that satisfy different resource requirements.
\item We present TDNN-NAS, an efficient search algorithm to boost the efficiency of the supernet: The progressive shrinking method requires no retraining and considerably reduces computation; The different-grained search space offers model efficiency balanced against accuracy.
\end{itemize}

The rest of this paper is organized as follows. In Section \ref{sec:works}, we discuss the related work in designing efficient speaker neural networks. The problem formulation of efficient architecture search are provided in Section \ref{sec:problem}. We present EfficientTDNN in Section \ref{sec:method}. Subsequently, the proposed method is evaluated on large-scale speaker datasets in Section \ref{sec:setting_results}. Sensitivity analysis is conducted to investigate multiple factors in Section \ref{sec:sensitivity}. We conclude this paper in Section \ref{sec:conclusion}.

\section{Related Work}\label{sec:works}

\subsection{Speaker Neural Architectures}
For speaker recognition, deep neural networks (DNNs) have successfully achieved state-of-the-art performance on challenging benchmarks, including SITW \cite{McLaren2016d}, SRE 2018 \cite{Sadjadi2019}, and VoxCeleb \cite{Nagrani2017,Chung2018,Nagrani2020}. 
Specifically, McLaren et al. \cite{McLaren2015} use DNN to extract bottleneck features to enhance robustness to microphone speech. Huang et al. \cite{Huang2018} apply a VGG-style CNN to learn embeddings via triplet loss. Wan et al. \cite{Wan2018} optimize LSTM by generalized end-to-end loss to improve speaker representations on text-dependent speaker verification. Snyder et al. \cite{Snyder2018} propose x-vector that maps variable-length utterances to fixed-dimensional embedding, which uses data augmentation to improve the performance of TDNN. Ramoji et al. \cite{Ramoji2019} extract neural embeddings from TDNN trained on a speaker discrimination task. Xie et al. \cite{Xie2019} use thin-ResNet with a GhostVLAD layer to investigate the effect of utterance length on the wild data. Garcia-Romero et al. \cite{Garcia-Romero2020} propose a wider ResNet by increasing channels in the early stages to achieve state-of-the-art performance. Desplanques et al. \cite{Desplanques2020} utilize several enhancements to TDNN, including residual connection, dense connection, and channel-dependent frame attention, which aggregate and propagate features of different hierarchical levels. Safari et al. \cite{Safari2020} alternatively employ a transformer encoder as a frame-level feature extractor to capture long-range dependencies. These works provide diverse promising temporal modeling architectures, which boost the performance of speaker embedding while remaining the same degree of parameters or computation.

\subsection{Neural Architecture Search}
The NAS technique provides a systematic methodology that designs neural architecture automatically. Zoph et al. \cite{Zoph2017} propose an RL-based approach to find an architecture that achieves state-of-the-art performance. However, such an approach would be computationally expensive, requiring thousands of different architectures trained from scratch. To avoid prohibitive computation, Bender et al. \cite{pmlr-v80-bender18a} propose a simple weight sharing for one-shot NAS. Liu et al. \cite{Liu2019} formulate the problem of NAS in a differentiable manner, which is orders of magnitude faster than non-differentiable techniques.
On the other hand, these works optimized on proxy tasks are not guaranteed optimal for the target task. To the end, Tan et al. combines training-aware NAS and scaling in a search space enriched Fused-MBConv to improve training speed and model parameters \cite{Tan2021}. Cai et al. \cite{Cai2019} propose to directly learn the architectures while handling hardware objectives via regularization loss. Subsequently, a once-for-all network \cite{Cai2020} is proposed to further reduce the computational cost by decoupling training and search, which provides a retraining-free approach and allows network pruning in a larger search space, i.e., depth, kernel size, and width.

In speaker recognition, AutoSpeech \cite{Ding2020} identifies the optimal operation combination in a neural cell and derives a CNN model by stacking the neural cell multiple times. Auto-Vector \cite{Qu2020} utilizes an evolutionary algorithm enhanced NAS method to discover a promising x-vector network. SpeechNAS \cite{Zhu2021} applies Bayesian optimization to conduct branch-wise and channel-wise selection in the search space of Densely connected TDNN. These works may suffer a weak correlation between the performance of the searched architectures and the ones trained from scratch \cite{Lu2020}. Besides, the requirement of retraining candidate networks causes linearly increasing computation, making the NAS not effectively scaled while handling various devices.

\section{Problem Formulation}\label{sec:problem}
OFA \cite{Cai2019} enables to quickly search for specialized architectures via decoupling the training and search, which significantly reduces the computational cost by getting rid of retraining. It motivates us to consider the NAS problem a two-stage problem consisting of supernet weights optimization and subnets search.

\subsection{Frequently Used Notations}
Frequently used notations and their meanings are given in Table \ref{tab:basic_notatino}, where 
$\mathcal{A}$, $a$, and $\mathbb{S}$ relate to architectures, $W_{\mathcal{A}}$ and $w_a$ are weights of architectures, 
$D$, $K$, and $C$ are variable dimensions, $v_i$ and $v_{i \to j}$ represent cells in the supernet.

\begin{table}[!t]
  \renewcommand{\arraystretch}{1.00}
  \caption{Frequently Used Notations}
  \label{tab:basic_notatino}
  \centering
  \begin{tabular}{ll}
  \toprule
  Notation & Description \\
  \midrule
  $\mathcal{A}$ & supernet that is nested in numerous $a$ \\
  $a$ & a subnet sampled from the supernet, $a\in\mathcal{A}$ \\
  $\mathbb{S}$ & subnet sampler used to sample a subnet, $a \leftarrow \mathbb{S}(\mathcal{A})$ \\
  $W_\mathcal{A}$ & weights of the supernet \\
  $w_a$ & weights of a network or subnet \\
  $D$ & depth of a network or a sequence of layers\\
  $K$ & kernel size of a convolution layer \\
  $C$ & width of a convolution or linear layer \\
  $v_i$ & a cell in an architecture \\
  $v_{i \to j}$ & a sequence of cells $v_i, v_{i+1}, \dots, v_{j}$, $j>i$ \\
  \bottomrule
  \end{tabular}
\end{table}

\subsection{Efficient Architecture Search}
The efficient architecture search for speaker recognition aims to find a speaker network that can efficiently extract speaker representation. Since training architectures from scratch is computationally prohibitive, the weight sharing technique is introduced. The coupled methods in that the supernet training and architecture search are coupled suffer from a weak ranking correlation \cite{Yu2020nas} probably caused by the optimization gap between the supernet and subnets \cite{Xie2020}. The decoupled one is thereby considered that the training and search are decoupled into two sequential stages to make real-world requirements feasible, such as parameters, multiply-accumulate operations (MACs), and latency. The efficient architecture search for speaker recognition can be formulated as a two-stage optimization problem.

\begin{enumerate}
  \item Supernet weights optimization is to minimize the loss function of the supernet on the training dataset $\mathcal{D}_\mathrm{train}$:
  \begin{equation}\label{eq:supernet_weight}
  W_{\mathcal{A}}=\mathop{\arg\min}_{W_{\mathcal{A}}}\mathcal{L}_{\mathrm{train}}(\mathcal{A},W_{\mathcal{A}})
  \end{equation}
  where supernet $\mathcal{A}$ contains all searchable architectures as subnets $a$, and all $a$ inherit weights directly from $\mathcal{A}$. 
  \item Subnets search is to find an architecture that satisfies the specific requirement on the evaluation dataset $\mathcal{D}_\mathrm{eval}$:
  \begin{equation}\label{eq:arch_search}
  a^*=\mathop{\arg\min}_{a\in\mathcal{A}}\mathrm{Metric}_{\mathrm{eval}}(a;W_{\mathcal{A}}(a))
  \end{equation}
  where $\mathrm{Metric}_{\mathrm{eval}}(\cdot)$ can be diverse, such as error rate, parameters, MACs, and latency.
\end{enumerate}

\section{Efficient Architecture Search Method}\label{sec:method}
To address the problem of the efficient architecture search for speaker recognition, we propose EfficientTDNN that consists of a TDNN-based supernet and the TDNN-NAS algorithm. The proposed supernet employs temporal convolution and feature aggregation, allowing scaling networks in both size and connection. The TDNN-NAS algorithm enables to search for TDNN-based subnets under the specific constraint.

\subsection{Supernet Design}\label{subsec:search_space}
Supernet determines the search space of the speaker network and plays an essential role in improving the performance of speaker representation. The TDNN-based supernet is proposed for speaker recognition, and several enhancement techniques are applied to further improve speaker representation performance. There are two steps to design a supernet: the macro and micro architectures of the space. The macro architecture determines the overall backbone of the extractor, such as TDNN architecture, while the micro architectures named cells determine the details of each network unit, such as temporal convolution and attentive feature aggregation.

\begin{figure}[!t]
  \centering
  \includegraphics[width=8.6cm]{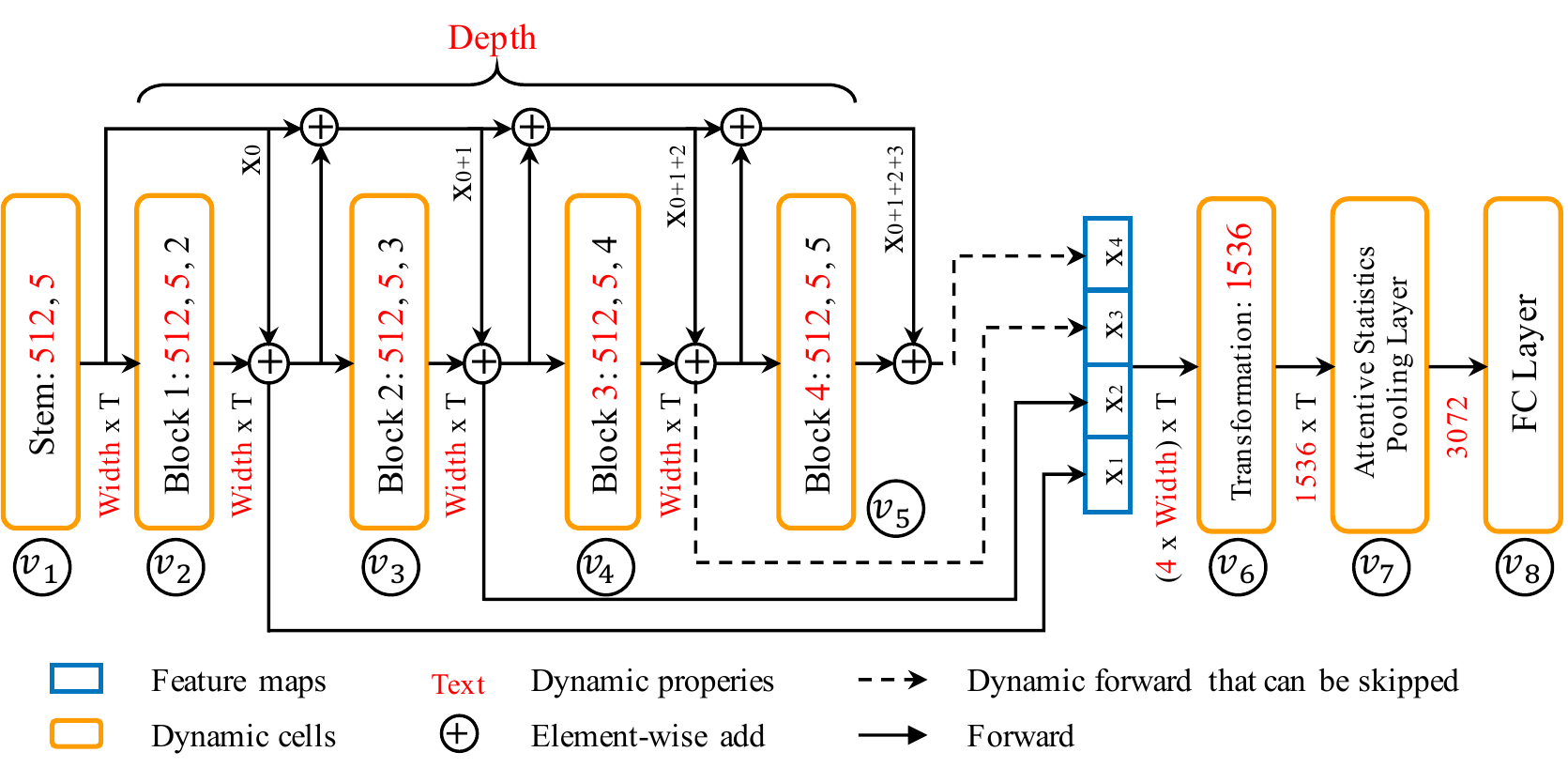}
  \caption{The proposed TDNN-based supernet of EfficientTDNN. The supernet is shallow with eight cells, while it is wide with up to 512 channels in $v_{1 \to 5}$ and 1,536 features in $v_{6 \to 8}$.}
  \label{fig:supernet_arch}
\end{figure}

\begin{definition}[Supernet]\label{def:supernet}
  Supernet is an over-parameterized neural network that contains dynamic components and enables sampling subnets with different architectures.
\end{definition}

The supernet of EfficientTDNN is built illustrated in Figure \ref{fig:supernet_arch}, which employs several architectural enhancement techniques from the design of ECAPA-TDNN \cite{Desplanques2020}, such as residual connections \cite{He2016}, dense connections \cite{Huang2017}, squeeze-and-excitation (SE) blocks \cite{Hu2020}, and Res2Net blocks \cite{Gao2021}. The proposed supernet is summarized as follows.

\begin{enumerate}
  \item $v_1$: a stem of input supports dynamic kernel and width.
  \item $v_{2 \to 5}$: a stack of blocks that contain Res2Net and SE support dynamic depth, kernel, and width.
  \item $v_6$: a transformation of concatenated features supports dynamic width.
  \item $v_7$: an attentive statistics pooling layer has dynamic width determined by $v_6$ output.
  \item $v_8$: a fully connected layer (FC) has dynamic width determined by $v_7$ output.
\end{enumerate}

Accordingly, cells with the dynamic depth are $v_{2 \to 5}$, cells with the dynamic kernel are $v_{1 \to 5}$, and cells with the dynamic width are $v_{1 \to 6}$. On the other hand, $v_7$ and $v_8$ are determined by $v_6$ width, which means they are constrained.

Compared to the original ECAPA-TDNN \cite{Desplanques2020}, there are mainly two differences: (1) Two blocks of $v_4$ and $v_5$ can be skipped; (2) Kernels of $v_{1 \to 5}$ are expanded to 5; (3) Width of $v_{1 \to 8}$ varies and weight-sharing.

The supernet supports dynamic depth, kernel, and width. Specifically, the depth is the number of blocks, where $D \in \{2, 3, 4\}$. It reduces as the last $d$ blocks are skipped, where $d \in \{0, 1, 2\}$. The kernel is the kernel size of the convolution layers at $v_1$ and the middle layer at $v_{1 \to 5}$, $K \in \{1, 3, 5\}$. The width is the number of channels or features in $v_{1 \to 8}$, where $C \in \{C_\mathrm{min}, C_\mathrm{min} + c, \dots, C_\mathrm{max} - c,  C_\mathrm{max}\}$, $C_\mathrm{min}$ is the given minimum width, $C_\mathrm{max}$ is the maximum width, and $c$ is an increasing step, which determines the density of sampled networks. We give that $C_\mathrm{min}=128$ and $C_\mathrm{max}=512$ for $v_{1 \to 5}$ as well as $C_\mathrm{min}=384$ and $C_\mathrm{max}=1536$ for $v_{6 \to 8}$.

\begin{definition}[Degrees of freedom]\label{def:spacefreedom}
  Degrees of freedom are the number of independent dimensions that are allowed to be variable in the formal description of the architecture of the supernet.
\end{definition}

According to Definition \ref{def:spacefreedom}, the proposed supernet has $1+5+6=12$ degrees of freedom, i.e., the depth has 1 degree, the kernel has 5 degrees, and the width 6 degrees.

\begin{definition}[Subnet]\label{def:subnetwork}
  Subnet is a neural network that is sampled from the supernet and can be encoded as a 3-tuple of elements of degrees of freedom of the network, $a\equiv (D, \mathbf{K}, \mathbf{C}) \in \mathcal{A}$, where $\mathbf{K}=\{K_i\}_{i=1}^{D+1}$, $\mathbf{C}=\{C_i\}_{i=1}^{D+2}$, and the subscript $i$ of $K$ and $C$ is the number of cells.
\end{definition}

For example, $(3, \{3\}_{i=1}^{4}, \{256\}_{i=1}^{4} \cup \{768\})$ denotes a subnet of 3 blocks with kernel sizes of 3, where $v_5$ is skipped. The width of $v_{1 \to 5}$ are 256, while $v_{6 \to 8}$ are 768.
The bound of the sampled architectures can be summarized: the largest subnet sampled from the supernet is $(4, \{5\}_{i=1}^{5}, \{512\}_{i=1}^{5} \cup \{1536\})$, while the smallest one is $(2, \{1\}_{i=1}^{3}, \{128\}_{i=1}^{3} \cup \{384\})$.

We consider the number of subnets as the size of space.

\begin{definition}[Sizes of space]\label{def:sizesofspace}
  Sizes of space are the number of independent subnets that are derived from a supernet.
\end{definition}

For example, the variable depth, kernel, and width lead to $145 \times ((3 \times 49)^3 + (3 \times 49)^4 + (3 \times 49)^5) \approx 1.0 \times 10^{13}$ subnets in a space of $c=8$ while $10 \times ((3 \times 4)^3 + (3 \times 4)^4 + (3 \times 4)^5) \approx 2.7 \times 10^{6}$ subnets in a space of $c=128$.

\begin{figure}[!t]
  \centering
  \includegraphics[width=8cm]{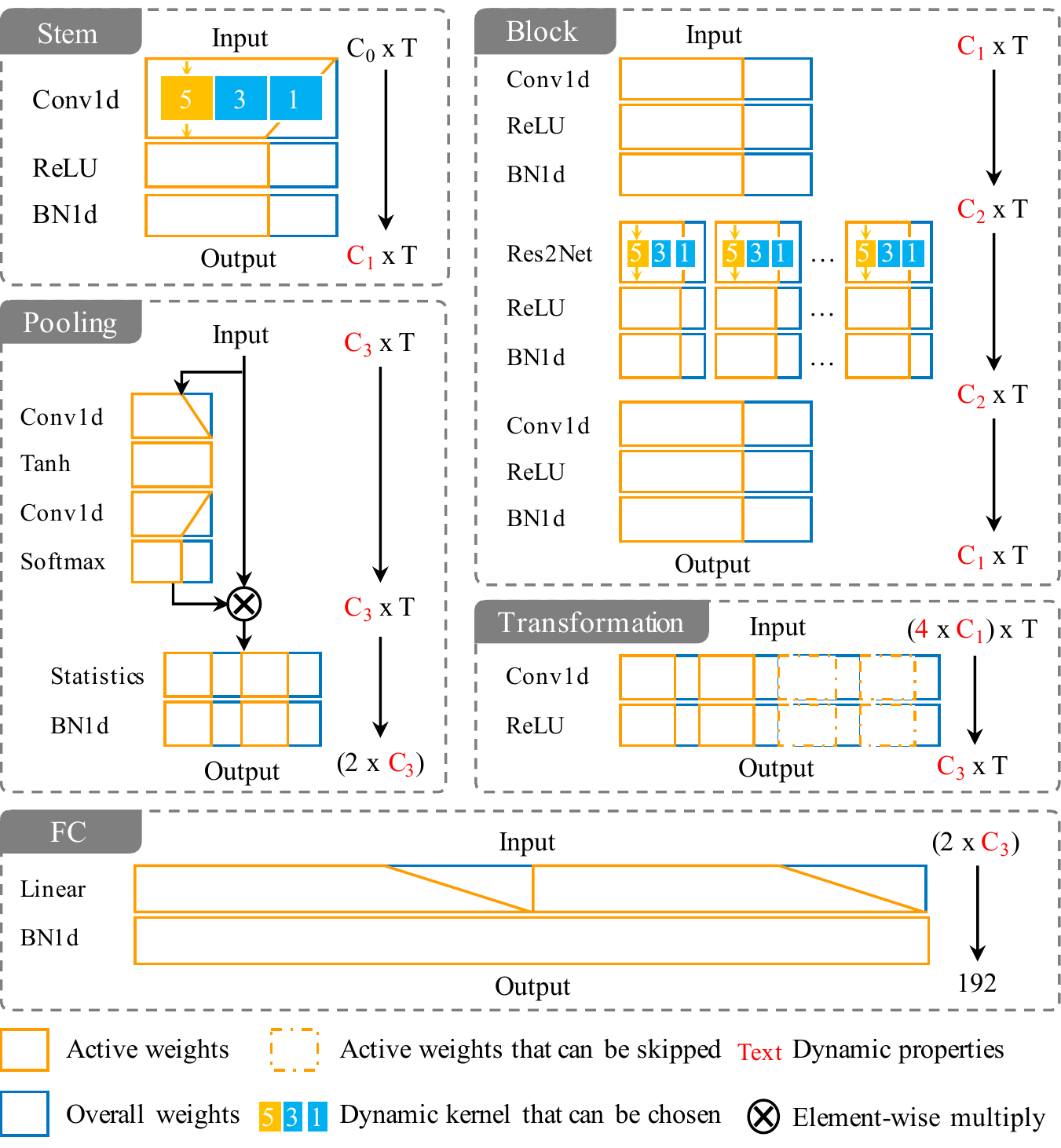}
  \caption{The details of micro architectures in the proposed EfficientTDNN, such stem, blocks, transformation, pooling, and FC. The orange and blue boxes are active and overall weights, respectively. The red text points to the variable input and output size at a cell and bridges the forward path.}
  \label{fig:dynamic_resblock}
\end{figure}

Figure \ref{fig:dynamic_resblock} illustrates the details of dynamic cells. It is clear to demonstrate how a subnet or a forward path is generated from the supernet. There are five different types of cells, i.e., \emph{stem}, \emph{block}, \emph{transformation}, \emph{pooling}, and \emph{FC}.

\begin{enumerate}
  \item \textbf{Stem} $v_1$ is a 1-dimensional convolution (Conv1d) followed by ReLU activation and 1-dimensional batchnorm (BN1d), as Conv1dReLUBN. The size of the input is fixed with $C_0$ channels and $T$ frames, i.e., $C_0 \times T$. The input is first expanded to $C_1$ channels and is applied to ReLU and BN1d. The output is generated with the size of $C_1 \times T$, where $C_1 \in [128, 512]$. The kernel size is dynamic and is chosen from $\{1, 3, 5\}$.
  \item \textbf{Block} $v_{2 \to 5}$ consists of three parts, where the first and last parts are Conv1dReLUBN with the kernel sizes of 1, and the middle part is a dilated convolution layer with a Res2Net module followed by ReLU and BN1d. The first part expands the input of $C_1 \times T$ to $C_2 \times T$, where $C_2 \in [128, 512]$. There can be four different $C_2$ as four blocks in the supernet. The output from Res2Net has the same size as the input, and then it is reduced to $C_1 \times T$ after the last part. Thus, the sizes of input and output are equal. The dynamic kernel of the Res2Net is done as similar as that of $v_1$.
  \item \textbf{Transformation} $v_6$ is a Conv1d followed by ReLU. The splices of the output of $v_{2 \to 5}$ are concatenated and serve as the input of this cell. The input is expanded to $C_3 \times T$, where $C_3 \in [384, 1536]$.
  \item \textbf{Pooling} $v_7$ contains an attention and a temporal statistics followed by BN1d. The attention calculates the importance of temporal features via a Conv1d with the kernel size of 1. The channel-wise weighted statistics are then calculated as a concatenated vector of mean and standard deviation, which convert the input from $C_3 \times T$ to $2C_3$.
  \item \textbf{FC} $v_8$ consists of a linear layer and a BN1d, which extracts an embedding that represents the speaker.
\end{enumerate}

\begin{figure}[!t]
  \centering
  \includegraphics[width=8cm]{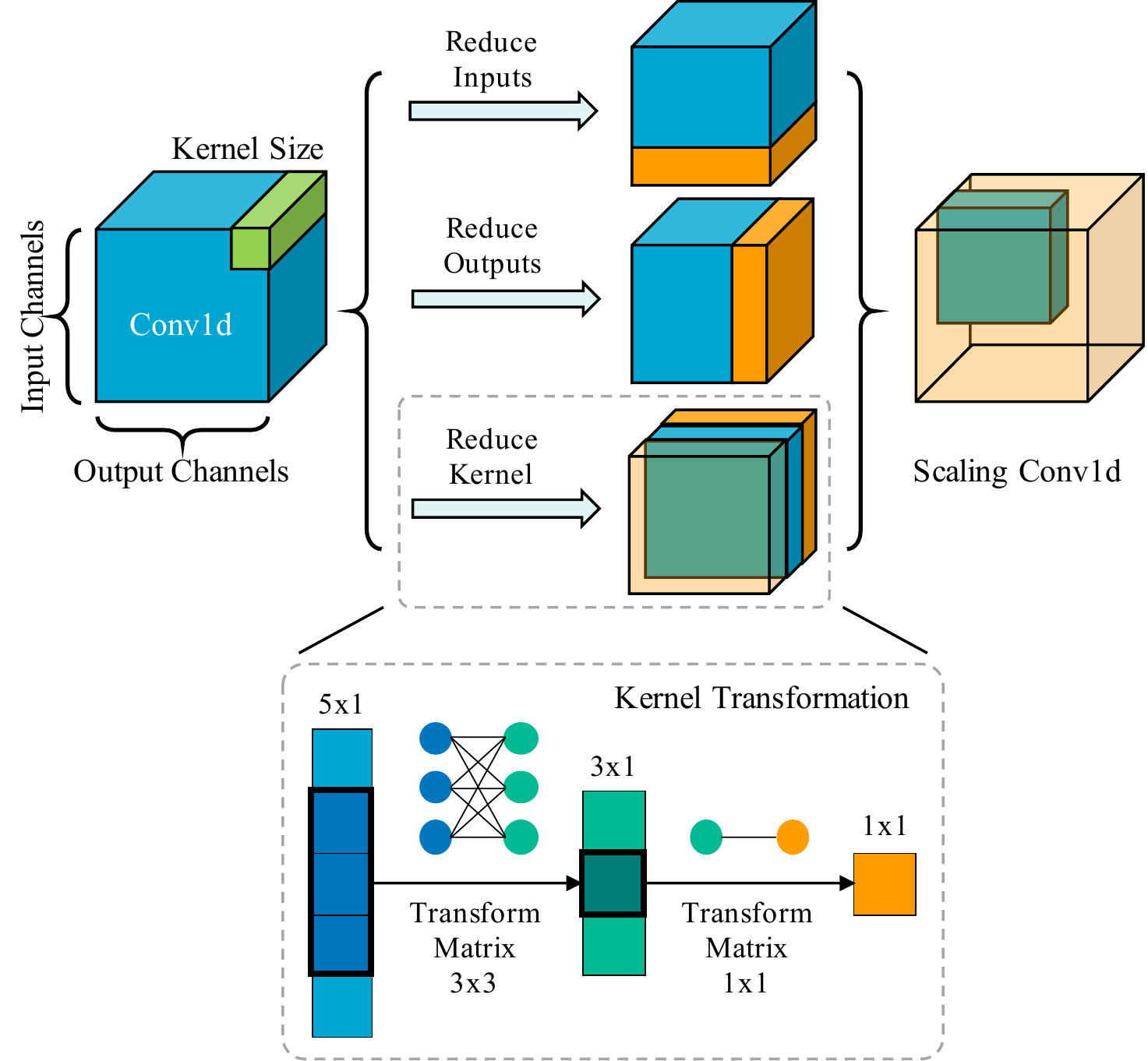}
  \caption{Scaling dynamic Conv1d in input channels, output channels, and kernel size. Kernel transformation matrices are applied to dynamic kernels.}
  \label{fig:dynamic_kernel}
\end{figure}

These dynamic components can be decomposed into three basic operations: dynamic Conv1d, BN1d, and linear layer.
These dynamic operations are the same as the traditional ones but perform with active weights as shown in Figure \ref{fig:dynamic_kernel}.
The weights of dynamic Conv1d at the $i$-th input channel and the $j$-th output channel are derived from the overall weights $W$ with active input and output channels indexes.
Likewise, the dynamic BN1d calculates mini-batch mean and variance with active channels.
The dynamic linear layer ignores inactive weights associated with dropped inputs.

However, sampling subnets from the supernet is challenging because the weights of dynamic components are shared and required to adapt to each other. For example, the weights of the centering sub kernel serve the subnets with different kernel sizes. 
Forcing these weights same probably degrades the performance of subnets with different depth, kernel, or width. 
Inspired by \cite{Cai2020}, a dynamic kernel of Conv1d is proposed to address this problem. As shown in Figure \ref{fig:dynamic_kernel}, when reducing the kernel size, a linear transformation is applied to the center of a larger kernel to create a weighted smaller kernel, e.g., the center of a kernel 5 serves a kernel 3. Kernel transformation matrices are employed as (\ref{eq:dynamic_kernel:1}) and (\ref{eq:dynamic_kernel:2}). These separate matrices are used for two sizes, i.e., $W_K^1 \in \mathbb{R}^{3 \times 3}$ is applied to kernel 5, while $W_K^2 \in \mathbb{R}^{1 \times 1}$ is applied to kernel 3. There are extra parameters in the supernet in $v_{1 \to 5}$, which is negligible compared to the whole architecture. Also, these parameters are eliminated when a subnet is derived from the supernet by performing kernel transformations.
\begin{IEEEeqnarray}{rcl}
  W_{\mathrm{kernel}}^{1}~&=&~W_K^1\mathrm{Center}(W_{\mathrm{kernel}})\IEEEyesnumber\label{eq:dynamic_kernel:1}\\
  W_{\mathrm{kernel}}^{2}~&=&~W_K^2\mathrm{Center}(W_{\mathrm{kernel}}^{1})\label{eq:dynamic_kernel:2}
\end{IEEEeqnarray}
where $W_{\mathrm{kernel}}\in\mathbb{R}^{5 \times 1}$, $W_{\mathrm{kernel}}^1\in\mathbb{R}^{3 \times 1}$, and $W_{\mathrm{kernel}}^2\in\mathbb{R}^{1 \times 1}$ denote the weights of kernel sizes of 5, 3, and 1, respectively. $W_K^1$ and $W_K^2$ are initialized in the form of the identity matrix.

The inference time complexity of a subnet $O(a)$ can be summarized as Theorem \ref{thm:time_inference}, regardless of the number of nonzero weights and nonzero MACs that reflect the theoretical minimum requirements of storage and computation \cite{Sze2017}.

\begin{theorem}[Bound of inference time]\label{thm:time_inference}
  Let $a\in\mathcal{A}$, then the inference time of a subnet with the minimum depth, the minimum kernel size, and the minimum width is a lower bound of $O(a)$, while the inference time of a subnet with the maximum depth, the maximum kernel size, and the maximum width is an upper bound of $O(a)$ so that $\forall~a \in \mathcal{A}$ we have
  \begin{IEEEeqnarray}{c}
    O(a_\mathrm{min}) \leq O(a) \leq O(a_\mathrm{max})\IEEEnonumber
  \end{IEEEeqnarray}
  where 
  \begin{IEEEeqnarray}{rcl}
    a_\mathrm{min}~&\equiv&~(D_\mathrm{min}, \mathbf{K}_\mathrm{min}, \mathbf{C}_\mathrm{min})\IEEEnonumber\\
    a_\mathrm{max}~&\equiv&~(D_\mathrm{max}, \mathbf{K}_\mathrm{max}, \mathbf{C}_\mathrm{max})\IEEEnonumber
  \end{IEEEeqnarray}
\end{theorem}

\begin{IEEEproof}
  By Definition \ref{def:subnetwork}, the minimum and maximum subnets are $(D_\mathrm{min}, \mathbf{K}_\mathrm{min}, \mathbf{C}_\mathrm{min})$ and $(D_\mathrm{max}, \mathbf{K}_\mathrm{max}, \mathbf{C}_\mathrm{max})$, respectively, $a_\mathrm{min}$ and $a_\mathrm{max}$ for clarity.
  \begin{IEEEeqnarray}{rcl}
    a_\mathrm{min}~&=&~(2, \{1, 1, 1\},       \{128, 128, 128, 384\})\IEEEnonumber\\
    a_\mathrm{max}~&=&~(4, \{5, 5, 5, 5, 5\}, \{512, 512, 512, 512, 512, 1536\})\IEEEnonumber
  \end{IEEEeqnarray}
  It clearly indicates that $\exists~a_1, a_2 \in \mathcal{A}$ we have
  \begin{IEEEeqnarray}{c}
    O(a_1) = O(a_\mathrm{min})\IEEEnonumber\\
    O(a_2) = O(a_\mathrm{max})\IEEEnonumber
  \end{IEEEeqnarray}

  Let $a \in \mathcal{A}$, the inference time of $a$ comes from static units and dynamic components as Definition \ref{def:supernet} and \ref{def:subnetwork}. The inference time of those static units is constant, but the inference time of dynamic components is affected by dynamic transformations.

  As the transformations of (\ref{eq:dynamic_kernel:1}) and (\ref{eq:dynamic_kernel:2}), $\forall~a \in \mathcal{A}$, for $v_{1 \to 5}$ we have 
  \begin{IEEEeqnarray}{rcl}
  W_{\mathrm{kernel}}(a)~&\in&~\{W_{\mathrm{kernel}}^2, W_{\mathrm{kernel}}^1, W_{\mathrm{kernel}}\}\IEEEnonumber
  \end{IEEEeqnarray}

  By the formulation of the dynamic Conv1d, the inference time increases as the kernel size enlarges. This implies that for $v_{1 \to 5}$ we have
  \begin{IEEEeqnarray}{ccccc}
    O(W_{\mathrm{kernel}}^2)~&<&~O(W_{\mathrm{kernel}}^1)~&<&~O(W_{\mathrm{kernel}})\IEEEnonumber
  \end{IEEEeqnarray}

  The dynamic components of $a_\mathrm{max}$ and $a_\mathrm{min}$ involve $W_\mathrm{kernel}$ and $W_\mathrm{kernel}^2$, respectively, for $v_{1 \to 5}$. Furthermore, the inference time of dynamic BN1d and linear layer increase as the size of input and output enlarge. Thus $\forall a \in \mathcal{A}$ we conclude

  \begin{IEEEeqnarray}{c}
    O(a_\mathrm{min}) \leq O(a) \leq O(a_\mathrm{max})\IEEEnonumber
  \end{IEEEeqnarray}
\end{IEEEproof}

Therefore, the minimum and maximum efficiency can be estimated by performing inference on $a_\mathrm{min}$ and $a_\mathrm{max}$.

\subsection{Architecture Search}\label{subsec:weights_optimization}

\begin{algorithm}
  \caption{Progressive Training Method}\label{alg:weight_optimization}

  \SetAlgoCaptionSeparator{.}
  \SetAlgoLined
  \DontPrintSemicolon
  \SetNoFillComment
  \SetSideCommentLeft
  \SetKwComment{Comment}{$\triangleright$\ }{}
  \SetKw{KwTo}{in}
  
  \KwIn{supernet $\mathcal{A}$, weights $W_\mathcal{A}$, subnet sampler $\mathbb{S}$, loss function $\mathcal{L}$, training data $\mathcal{D}_{\mathrm{train}}$}
  \KwOut{the trained weights of the supernet $W_\mathcal{A}$}

  Set sampling space $s \leftarrow \emptyset$ for $\mathcal{A}$\;
  
  Initialize the weights of the supernet $W_\mathcal{A}$\;
  
  \For{\emph{task \KwTo} \{largest, kernel, depth, width\}}{
    \uIf{\emph{task \textbf{is}} largest}{
      $\mathcal{A} \leftarrow $ set $\mathcal{A}$ as the largest architecture \;
    }
    \uElseIf{\emph{task \textbf{is}} kernel}{
      $s \leftarrow s \cup \{\mathrm{kernel}: \{1, 3, 5\}\}$ \;
    }
    \uElseIf{\emph{task \textbf{is}} depth}{
      $s \leftarrow s \cup \{\mathrm{depth}: \{2, 3, 4\}\}$ \;
    }
    \Else{
      \For{\emph{phase \KwTo \{1, 2\}}}{
          \uIf{\emph{phase \textbf{is} 1}}{
            $s \leftarrow s \cup \{\mathrm{width}: \{0.5, 0.75, 1\}\}$ \;
          }
          \Else{
            $s \leftarrow s \cup \{\mathrm{width}: \{0.25, 0.35\}\}$\;
          }
        }
    }
    $W_\mathcal{A} \leftarrow $ update $W_\mathcal{A}$ using Algorithm \ref{alg:dynamic_training} \;
  }
  \KwRet{$W_\mathcal{A}$} \;
\end{algorithm}

The supernet comprises various subnets of different sizes, where small subnets are nested in large subnets. The TDNN-NAS algorithm is proposed to enable efficient architecture search from the TDNN-based supernet. 
The proposed algorithm consists of the progressive training method and constrained search algorithm. The progressive training method ensures that the sampled subnets extract the speaker embedding effectively without retraining. The constrained search algorithm searches for the specialized subnet efficiently that satisfies the given budget.

\begin{figure}[!t]
  \centering
  \includegraphics[width=8cm]{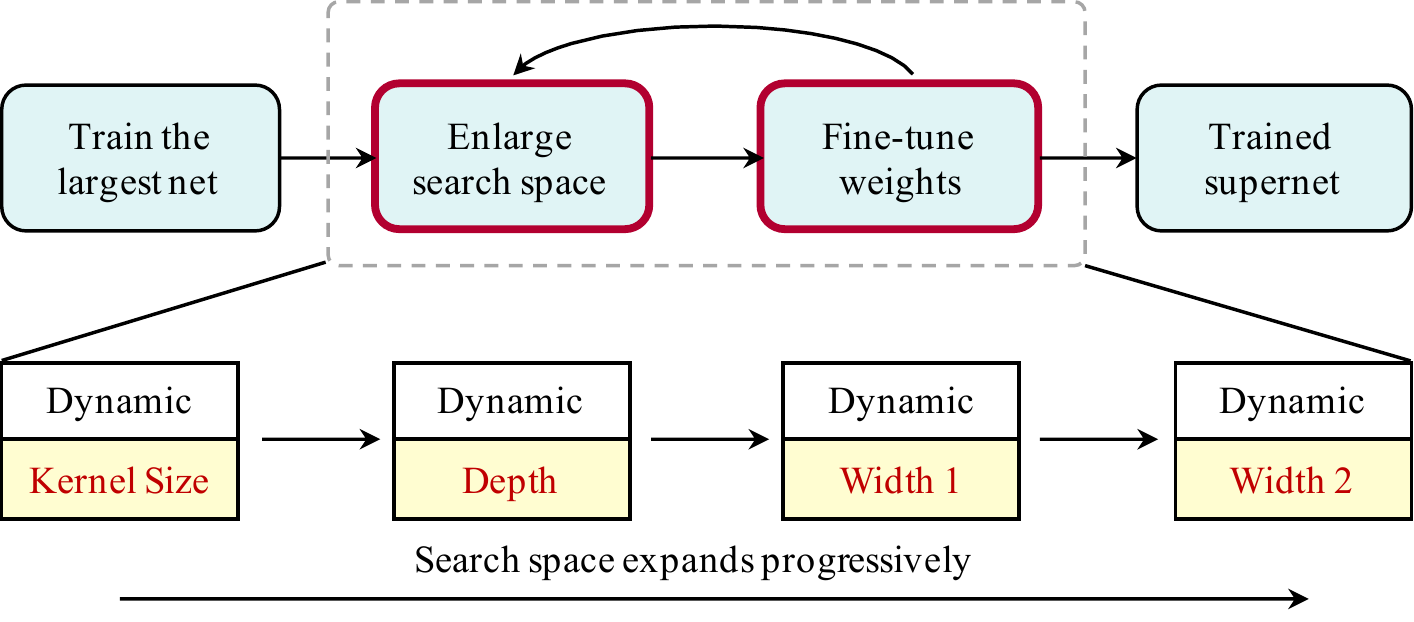}
  \caption{The process of the progressive training method.}
  \label{fig:progressive_train}
\end{figure}

The progressive training method is proposed based on the TDNN-based supernet with dynamic depth, kernel, and with as illustrated in Figure \ref{fig:progressive_train}, which is helpful to mitigate interference between subnets and overcome the challenge of optimizing a huge search space. Specifically, the weights optimization problem (\ref{eq:supernet_weight}) is decomposed into five stages as follows.
\begin{IEEEeqnarray}{l}\label{eq:progressive_optimization}
  \begin{IEEEeqnarraybox}[][c]{rl}\label{eq:progressive_dynamic}
    W_{\mathcal{A}^-}=\mathop{\arg\min}_{W_{\mathcal{A}^-}}\mathcal{L}_{\mathrm{train}}(\mathcal{A}^-, W^-)
  \end{IEEEeqnarraybox}\IEEEyesnumber
\end{IEEEeqnarray}
where $\mathcal{A}^-$ is a variant of the search space with the specific sampling dimensions, and $W^-$ represents the weights. For example, $\mathcal{A}^{\mathrm{kernel}}$ is a supernet with variable kernel sizes with the weights $W^{\mathrm{kernel}}$. 

\begin{algorithm}
  \caption{Dynamic Path Training}\label{alg:dynamic_training}

  \SetAlgoCaptionSeparator{.}
  \SetAlgoLined
  \DontPrintSemicolon
  \SetNoFillComment
  \SetSideCommentLeft
  \SetKwComment{Comment}{$\triangleright$\ }{}
  \SetKwProg{Def}{def}{:}{end}

  \KwIn{weights $W_\mathcal{A}$, subnet sampler $\mathbb{S}$, sampling space $s$, forward paths $M$, loss function $\mathcal{L}$, training data $\mathcal{D}_{\mathrm{train}}$, epochs $T$, data augmentation $\mathcal{F}_\mathrm{aug}$}
  \KwOut{the trained weights $W_\mathcal{A}$}

  \Def{\emph{DynamicTrain($W_\mathcal{A}$, $\mathbb{S}$, $s$, $M$, $\mathcal{L}$, $\mathcal{D}_{\mathrm{train}}$, $T$, $\mathcal{F}_\mathrm{aug}$)}}{
    $\mathbb{S} \leftarrow$ update $\mathbb{S}$ by using $s$ \;
    
    $t \leftarrow 0$ \;
    
    \While{$t < T$}{
      \For{\emph{batch$_i$ \textbf{in} $\mathcal{D}_{\mathrm{train}}$}}{
        $f_\mathrm{aug} \leftarrow $ choose from $\mathcal{F}_\mathrm{aug}$ randomly \;
        
        $\mathrm{batch}_i \leftarrow f_\mathrm{aug}(\mathrm{batch}_i)$ \;
        
        $\nabla = \emptyset$ \;
        
        \For{$i=1$ \KwTo $M$}{
          $a \leftarrow \mathbb{S}(\mathcal{A})$ \;
          
          $w \leftarrow W_\mathcal{A}(a)$ \; 
          
          $\mathcal{L} \leftarrow $ cross-entropy on batch$_i$ \;
          
          $\nabla \leftarrow \nabla \cup \partial \mathcal{L} / \partial w$ \;
        }
        $W_\mathcal{A} \leftarrow $ update $W_\mathcal{A}$ using Adam with $\nabla$ \; 
      }
      $t \leftarrow t + 1$ \;
    }
    \Return{$W_\mathcal{A}$} \;
  }
\end{algorithm}

\begin{table}[!t]
  \renewcommand{\arraystretch}{1.00}
  \caption{Sizes of Space in Training Tasks}
  \label{tab:train_space}
  \centering
  \begin{tabular}{llr}
  \toprule
  Stage & Dynamic Dimension & Sizes of Space \\
  \midrule
  \emph{largest} & $\emptyset$ & 1 \\
  \emph{kernel} & $~ \cup \{\mathrm{kernel}: \{1, 3, 5\}$ & 243 \\
  \emph{depth} & $~ \cup \{\mathrm{depth}: \{2, 3, 4\}$ & 351 \\ 
  \emph{width 1} & $~ \cup \{\mathrm{width}: \{0.5, 0.75, 1\}$ & 199,017 \\
  \emph{width 2} & $~ \cup \{\mathrm{width}: \{0.25, 0.35\}$ & 4,066,875 \\
  \bottomrule
  \end{tabular}
\end{table}

A variant of the single path one-shot routing approach \cite{guo2020single} is proposed to solve these subproblems sequentially as shown in Algorithm \ref{alg:weight_optimization} and \ref{alg:dynamic_training}. 
As shown in Figure \ref{fig:progressive_train}, the proposed training method enforces training from large subnets to small subnets in a progressive manner to overcome the limitation of dramatically enlarging search space makes the weights of cells hard to adapt to each other. 
First, the largest network is optimized with the maximum kernel, depth, and width. Next, the supernet is fine-tuned progressively to support smaller subnets by gradually expanding the search space. Specifically, after the largest architecture is trained, the dynamic kernel is supported and is chosen from $\{1, 3, 5\}$ at $v_{1 \to 5}$, while the depth and width keep the maximum value. Subsequently, the dynamic depth and width are supported. Since the range of dynamic width is large, the search space of width is separate by taking half of the maximum value as the dividing line. According to Definition \ref{def:sizesofspace}, sizes of space at different stages are listed in Table \ref{tab:train_space}, which reveals a significant increase in sizes of space after \emph{depth}.

As shown in Algorithm \ref{alg:weight_optimization}, the proposed training approach divides the search space into five sequential parts and separately optimizes those weights, i.e., \emph{largest}, \emph{kernel}, \emph{depth}, \emph{width 1}, and \emph{width 2}. In each stage, the subnets sampler $\mathbb{S}$ samples one or more forward paths randomly, which is employed to update the used parameters in the supernet.
Specifically, in the stage of \emph{largest}, the supernet retains the maximum architecture, where the weights except $W_K^1$ and $W_K^2$ are updated. Next, in \emph{kernel}, the forward path allows sampling subnets from multiple kernel sizes, e.g., $\{1,3,5\}$. As outlined in Algorithm \ref{alg:dynamic_training}, the weights of subnets are updated at each training step, including $W_K^1$ and $W_K^2$. In each step, $M$ subnets are applied to aggregate gradients, and the weights are updated along these forward paths. The stage of \emph{depth} is performed after the stage of \emph{kernel}. It makes the number of blocks within $[2,4]$. The \emph{width} is divided into two stages, where the width of $\{0.5, 0.75, 1.0\}$ and $\{0.25, 0.35\}$ of the maximum value are sequentially added into sampling space.

Once the supernet is trained, the constrained search algorithm is proposed to derive the specialized subnet for the given budget without retraining. The objective of the proposed search algorithm is to find the optimal architecture that optimizes the accuracy while satisfying the efficiency constraints as follows.
\begin{IEEEeqnarray}{l}\label{eq:search_acc}
  \begin{IEEEeqnarraybox}[][c]{rl}\label{eq:search_acc_efficiency}
  \mathop{\arg\max}_{a\in\mathcal{A}}&~\mathrm{ACC}_{\mathrm{val}}(a;W_{\mathcal{A}}(a))\\
  \mathrm{s.t.}&~\mathrm{Efficiency}(a)\leq\mathrm{Budget}
  \end{IEEEeqnarraybox}\IEEEyesnumber
\end{IEEEeqnarray}
where $\mathrm{ACC}_\mathrm{val}$ is an accuracy metric, $\mathrm{Efficiency}$ is an efficiency metric, and $\mathrm{Budget}$ is its budget. 

The design of sampling space and search strategy plays an essential role in solving search problems. Three modes of the search space are considered to investigate the impact of sampling space: the grid, coarse-grained, and fine-grained spaces. In the grid space, depth, kernel, and width are supposed as three independent integers. 
Specifically, the widths of $v_{1 \to 5}$ are the same, while the widths of $v_{6 \to 8}$ are three times as large as $v_{1 \to 5}$, e.g., $(4, \{3\}_{i=1}^{5}, \{256\}_{i=1}^{5} \cup \{768\})$. 
The exhaustive search is applied to such a small space, where architectures can be enumerated.
The coarse-grained space uses a sampling space used in the training process, where depth is chosen from $\{2,3,4\}$, kernel size is chosen from $\{1,3,5\}$, and width is chosen from $\{0.25, 0.35, 0.5, 0.75, 1.0\}$ of the maximum value. Compared to the coarse-grained space, the fine-grained space uses the width of $c=8$, which provides a subtle solution space.
For the large space, a random strategy is introduced as the baseline method. Then, a model predictive evolutionary algorithm (MPEA) is introduced as the advanced search algorithm based on the random search results, where a population of architectures is estimated in the given constraint, and the ones with a lower budget are retained.

MPEA employs the accuracy predictor and efficiency estimator to guide the direction of architecture search. The accuracy predictor and efficiency estimator are established for quickly estimating the accuracy and efficiency of an architecture. 
The accuracy predictor is built based on the pairs of accuracy metrics and one-hot subnet encoding. 
The efficiency estimator consists of a latency estimator, MACs counter, and parameters counter. 
The latency estimator is created via an operator-wise manner, e.g., measuring the stem with different widths and kernel sizes. 
The MACs and parameter counters are used to calculate MACs and parameters of subnets. 
Since the time cost by counting MACs and parameters with forwarding inference is non-trivial, a recursive manner that supports dynamic architectural hyperparameters is developed without inference. Specifically, the MACs and parameters are determined via dynamic Conv1d, BN1d, and linear layers.

\section{Experiments Settings and Results}\label{sec:setting_results}
\subsection{Datasets}\label{exp:datasets}
Experiments are conducted on the VoxCeleb datasets including VoxCeleb1 (Vox1) \cite{Nagrani2017} and VoxCeleb2 (Vox2) \cite{Chung2018}. The corpora are extracted from videos uploaded to YouTube as large-scale speaker recognition datasets collected in the wild. 
The VoxCeleb2 development set includes 1,092,009 utterances from 5,994 celebrities, and the VoxCeleb1-O (Vox1-O) test set includes 4,708 utterances from 40 speakers. There is no overlapping speaker between the development and test sets.

The feature is an 80-dimensional log Mel-filterbanks extracted from spectrograms within the given frequency limit of 20-7600 Hz. Pre-emphasis is first applied to the input signal using a coefficient of 0.97. Spectrograms are extracted with a hamming window of width 25 ms and step 10 ms with an FFT size of 512. The average and variance normalization is applied to the feature using instance normalization.

\begin{table}[!t]
  \renewcommand{\arraystretch}{1.00}
  \caption{Considered Options of Search Space}
  \label{tab:seach_space}
  \centering
  \begin{tabular}{lrr}
  \toprule
  Dimension & Options & \# Options \\
  \midrule
  Depth & 2, 3, 4 & 3 \\
  Kernel Size & 1, 3, 5 & 3 \\ 
  Coarse-grained Width $v_{1 \to 5}$ & $128, 176, 256, 384, 512$ & $5$ \\
  Coarse-grained Width $v_{6 \to 8}$ & $384, 536, 768, 1152, 1536$ & $5$ \\
  Fine-grained Width $v_{1 \to 5}$ & $128, 136, \dots, 512$ & $49$ \\
  Fine-grained Width $v_{6 \to 8}$ & $384, 392, \dots, 1536$ & $145$ \\
  \bottomrule
  \end{tabular}
\end{table}

\subsection{Supernet Preparation}\label{exp:supernet}
The considered options of the search space are given in Table \ref{tab:seach_space}. According to Theorem \ref{thm:time_inference}, the inference time of a subnet $O(a)$ is between $O(a_\mathrm{C2min})$ and $O(a_\mathrm{max})$.
\begin{IEEEeqnarray}{rcl}
  a_\mathrm{C2min}~&=&~(2, \{1\}_{i=1}^3, \{128, 128, 128, 384\})\IEEEnonumber\\
  a_\mathrm{max}~&=&~(4, \{5\}_{i=1}^5, \{512, 512, 512, 512, 512, 1536\})\IEEEnonumber
\end{IEEEeqnarray}

Different spaces lead to the different lower bound of inference time. For the training task of \emph{kernel}, \emph{depth}, and \emph{width 1}, their minimum inference time are derived from $a_\mathrm{Kmin}$, $a_\mathrm{Dmin}$, and $a_\mathrm{C1min}$, respectively,
\begin{IEEEeqnarray}{rcl}
  a_\mathrm{Kmin}~&=&~(4, \{1\}_{i=1}^5, \{512, 512, 512, 512, 512, 1536\})\IEEEnonumber\\
  a_\mathrm{Dmin}~&=&~(2, \{1\}_{i=1}^3, \{512, 512, 512, 1536\})\IEEEnonumber\\
  a_\mathrm{C1min}~&=&~(2, \{1\}_{i=1}^3, \{256, 256, 256, 768\})\IEEEnonumber
\end{IEEEeqnarray}

\subsection{Training Details}\label{exp:training}
Each stage in Algorithm \ref{alg:weight_optimization} is performed under AAM-Softmax \cite{Deng2019,Xiang2019} within 64 epochs using the 16-epoch cyclic learning rate between $10^{-8}$ and $10^{-3}$. The number of dynamic paths $M$ is set to $1$ with a uniform sampling strategy for all tasks except for \emph{largest}. The \emph{largest} task uses 2-second segments as input with data augmentation that consists of RIR dataset (reverb) \cite{Ko2017}, MUSAN dataset (music, speech, noise) \cite{Snyder2015}, open-source SoX effects (speedup, slowdown, compand) \cite{sox2021}, and SpecAugment (time masking, frequency masking) \cite{Park2019}. The other training tasks use 3-second segments with the data augmentation as the \emph{largest} task except SpecAugment.

\subsection{Evaluation Details}\label{exp:evaluating}
There are two different indexes to evaluate, i.e., accuracy and efficiency. Accuracy reports two metrics, namely, the equal error rate (EER) and the detection cost function (DCF$_{0.01}$). The former is the rate at which both acceptance and rejection errors are equal. The latter is adopted in NIST SRE 2018 \cite{Sadjadi2019} and VoxSRC 2019 \cite{Chung2019} given $C_\mathrm{target}=0.01$. Efficiency metrics contain CPU latency, CPU latency, MACs, and parameters, where CPU and GPU latency are estimated via their latency tables, and MACs are measured using a 3-second utterance. The latency table on Intel Xeon E5-2698 v4 CPU and NVIDIA Tesla V100 GPU are measured to estimate the latency of different architectures, where CPU latency uses a 3-second utterance and GPU latency uses 64 3-second utterances.

The scores for EER are calculated as similar as \cite{Heo2020} if there is no special mention. Specifically, two 4-second temporal segments are sampled at regular intervals from each test utterance. $2 \times 2 = 4$ cosine similarities are computed from each pair of segments, and their average is used as the score.
Before scoring, the mean and standard deviation of BN1d are calibrated via training utterances since changes of subnets lead to a mismatch of feature maps within layers. BN1d is reset using 6,000 3-second utterances in a batch of 32.

\subsection{Search Details}\label{exp:searching}
There are three different approaches, i.e., manual grid search, random search, and MPEA. The manual grid search constrains the search space as $(D, \{K\}_{i=1}^{D+1}, \{C\}_{i=1}^{D+1} \cup \{3C\})$, which contains 441 subnets. The random search considers two different-grained spaces, including the coarse-grained and fine-grained spaces, in which 10,000 subnets are sampled randomly. In MPEA, a three-layer feedforward neural network with 400 hidden units followed by ReLU activation in each layer is used as the accuracy predictor. The model is supervised by the mean absolute error between the normalized ground-true and predicted EER or DCF$_{0.01}$ on the Vox1-O test set. The MPEA is conducted as a constrained single objective using the Geatpy tool \cite{geatpy} with the population of 50, the mutation of 0.1, and the generation of 200.

\subsection{Speaker Verification}\label{exp:verification}

\begin{table}[!t]
  \renewcommand{\arraystretch}{1.00}
  \caption{Comparison of Accuracy on the Vox1-O Test between EfficientTDNN and State-of-the-Art CNN Models}
  \label{tab:sota_cnn}
  \centering
  \begin{tabular}{lrrcc}
  \toprule
  Sytems & MACs & Params & EER $\downarrow$ & DCF$_{0.01}$ $\downarrow$ \\
  \midrule
  D-TDNN$^{\star,1}$ \cite{Yu2020} & 14.93G & 2.36M &  1.81\% & 0.200 \\
  ResNet-20$^{\star,2}$ \cite{Hajibabaei2018} & 12.89G & 16.11M &  4.30\% & 0.413 \\
  Dual Attention$^{\star,2}$ \cite{Li2020} & 4.01G & 21.67M &  1.60\% & - \\
  H/ASP$^{\star,2}$ \cite{Heo2020} & 3.46G & 6.06M &  1.18\% & - \\
  ARET-25$^{1}$ \cite{Zhang2020} & 2.9G & 12.2M &  1.39\% & 0.199 \\
  ECAPA-TDNN$^{\star,1}$ \cite{Desplanques2020} & 1.45G & 5.79M &  1.01\% & 0.127 \\
  Fast ResNet-34$^{\star,2}$ \cite{Chung2020} & 672M & 1.40M & 2.22\% & - \\
  \midrule
  AutoSpeech$^{\star,3}$ \cite{Ding2020} & 3.63G & 15.11M &  8.95\% & - \\
  \midrule
  EfficientTDNN-Small & 204M & 0.90M & 2.20\% & 0.219 \\
  EfficientTDNN-Mobile & 571M & 2.42M & 1.41\% & 0.124 \\
  EfficientTDNN-Base & 1.45G & 5.79M & \textbf{0.94\%} & \textbf{0.089} \\
  \bottomrule
  \end{tabular}
\end{table}

\begin{figure*}[!t]
  \centering
  \includegraphics[width=16.6cm]{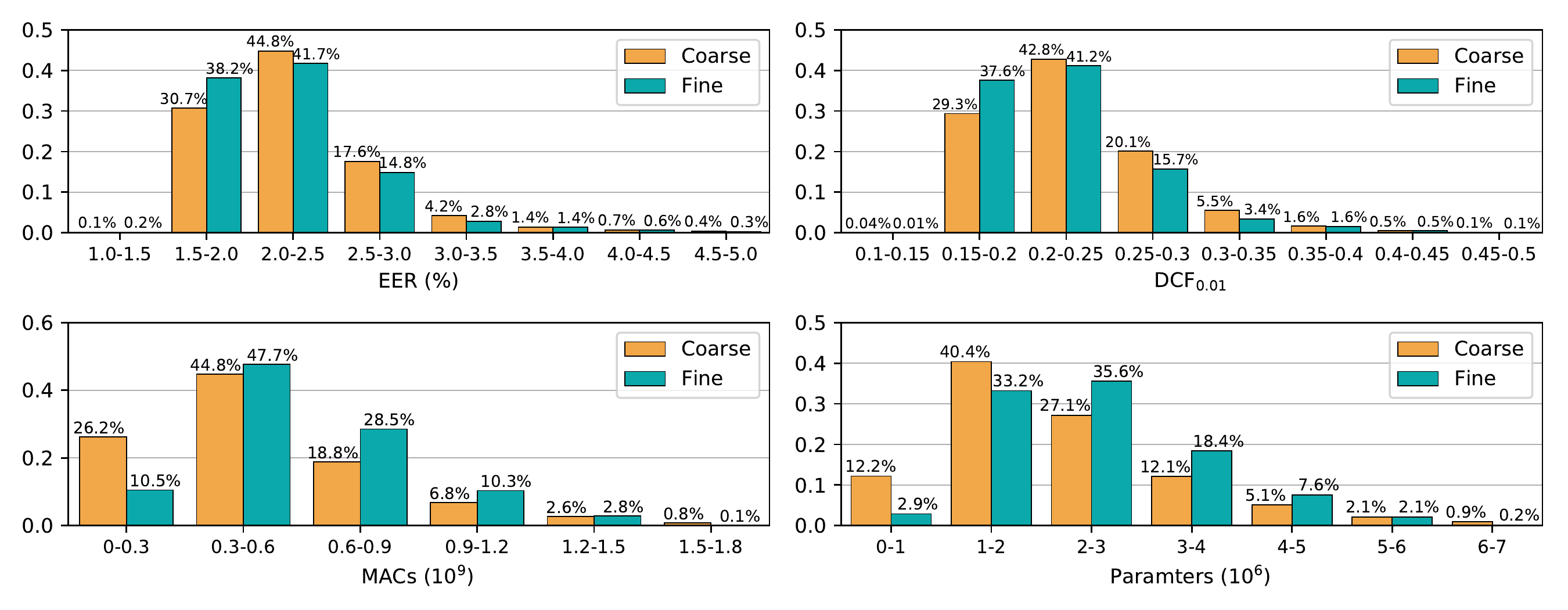}
  \caption{Distribution of accuracy and efficiency of two sets of 10,000 subnets randomly sampled from the coarse-grained and fine-grained spaces. The accuracy includes EER and DCF$_{0.01}$, and the efficiency includes MACs and parameters. The percentage of each bin is given for clarity.}
  \label{fig:grain_accuracy_efficiency}
\end{figure*}

Pruning a network reduces the requirements of storage and computation but degrades the accuracy. The accuracy of EfficientTDNN is first shown in order to study the effects of the network pruning. Table \ref{tab:sota_cnn} reports a comparison between EfficientTDNN and state-of-the-art CNN models on VoxCeleb1 Test, where $\star$ denotes that MACs and parameters are obtained by reproducing the model, and the superscripts of $^{1/2/3}$ represent TDNN-based, ResNet-based, and NAS-based architectures, respectively. Three subnets of EfficientTDNN are considered from different training stages of the supernet, i.e., the Small of $(2, \{3\}_{i=1}^{3}, \{256\}_{i=1}^{3} \cup \{400\})$ from \emph{width 2}, the Mobile of $(3, \{5, 3, 3, 3\}, \{384, 256, 256, 256, 768\})$ from \emph{width 1}, and the Base of $(3, \{5, 3, 3, 3\}, \{512\}_{i=1}^{4} \cup \{1536\})$ from \emph{depth}, respectively, which are produced via taking the whole utterance as inputs and applying top-300 adaptive score normalization based on the imposter cohort of the utterance-wise embeddings of 6,000 training utterances.

Compared with Fast ResNet-34, EfficientTDNN-Small attains competitive performance using 30\% MACs and 64\% parameters. It demonstrates the effectiveness of the small-size network sampled from the supernet trained after \emph{width 2}.

EfficientTDNN-Mobile from the supernet trained after \emph{width 1} achieves 1.41\% EER and 0.124 DCF$_{0.01}$ with 571M MACs in the mobile setting ($<$ 600M MACs). It outperforms D-TDNN, ResNet-20, Dual-Attention, and Fast ResNet-34 with less storage or computation. It suggests that the search space design provides a well-behaved supernet, creating an efficient architecture for resource-limited devices.

As the budget of storage and computation is neglected, Efficient-Base, the same architecture as ECAPA-TDNN, obtains 0.94\% EER and 0.089 DCF$_{0.01}$, which comes from the supernet trained after \emph{depth}. The architecture is superior to H/SAP, ARET, and ECAPA-TDNN, which suggests that the progressive training method concerning dynamic kernel and depth is helpful. It implies that the training in a dynamic architecture manner provides an alternative for improving speaker representation. 

Compared to the NAS method, AutoSpeech, EfficientTDNN achieves a significant improvement, which is a consequence of designing a search space more suitable for speaker recognition.

\begin{figure*}[!t]
  \centering
  \includegraphics[width=16.6cm]{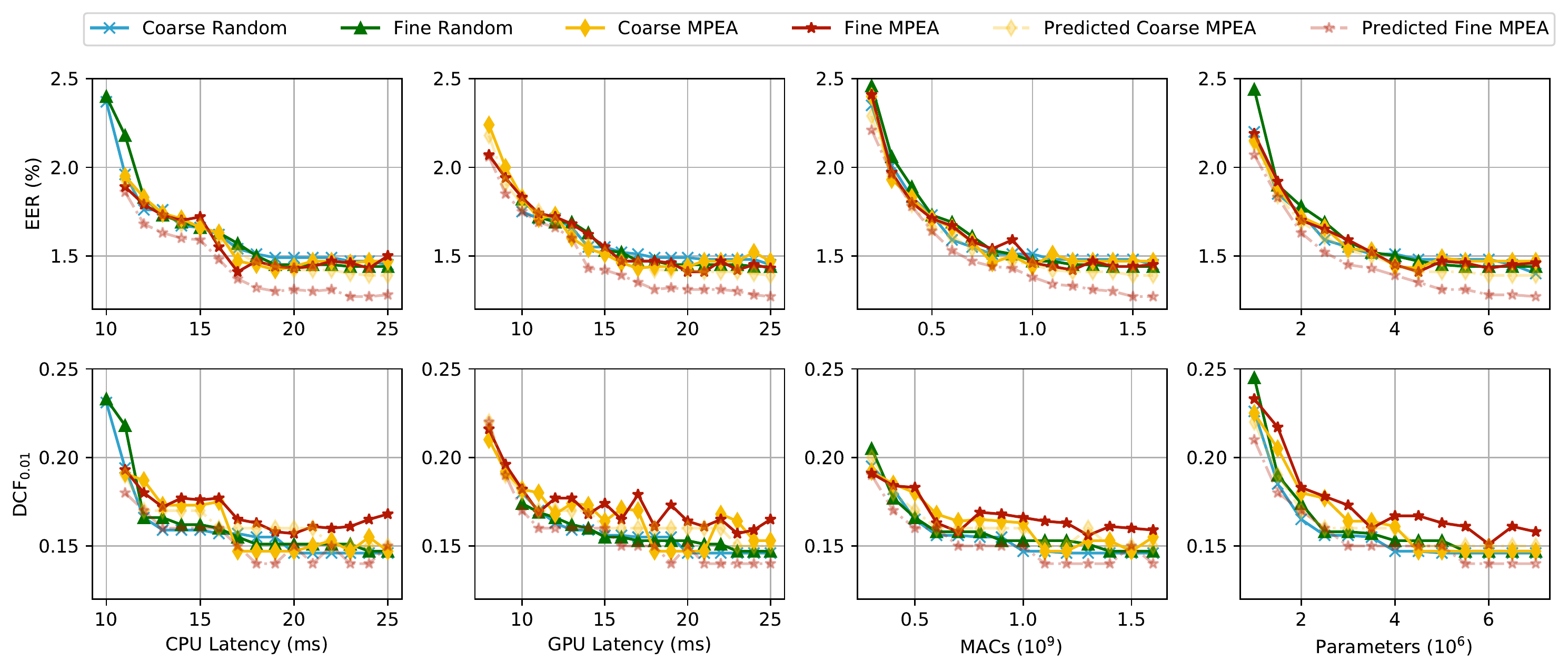}
  \caption{Trade-off between accuracy and efficiency with different search methods. `Coarse Random' denotes random search on a coarse-grained space, and `Coarse MPEA' conducts MPEA on it as the actual accuracy. In contrast, `Predicted Coarse MPEA' reports predicted accuracy from an accuracy predictor. `Fine Random,' `Fine MPEA,' and `Predicted Fine MPEA' are notations similar to `Coarse' methods. In each subfigure, the x-axis denotes the given constraints of an efficiency metric, and the y-axis denotes the accuracy of the found subnet.}
  \label{fig:tradeoff_accuracy_efficiency}
\end{figure*}

\subsection{Generalizing Different-grained Space}\label{exp:generalization}
Two sets of 10,000 subnets are sampled from different-grained spaces randomly to study the granularities of search space, and the distributions of their accuracy and efficiency are illustrated in Figure \ref{fig:grain_accuracy_efficiency}. It is intuitively problematic to achieve favorable accuracy on those architectures sampled from the fine-grained space that do not perform forward during the training process. However, in Figure \ref{fig:grain_accuracy_efficiency}, no subnet from the fine-grained space achieves the accuracy out of the range as those subnets from the coarse-grained space. It implies that the supernet trained progressively generalizes unseen cells.

Figure \ref{fig:grain_accuracy_efficiency} shows evident differences in EER, DCF$_{0.01}$, MACs, and parameters between the distributions of the coarse-grained and fine-grained spaces. The accuracy of subnets sampled from the fine-grained space emphasizes a higher accuracy, e.g., the percentage of the first and second bins is around 38\% and 37\% compared to around 31\% and 29\% from the coarse-grained spaces. It probably results from that subnets sampled from the fine-grained space highlight larger computation, such as MACs between 0.3G and 1.2G and parameters between 2M and 5M, as shown in the bottom subfigures of Figure \ref{fig:grain_accuracy_efficiency}. Searching for these candidates is helpful to find specialized architectures that maintain accuracy under computation-limited conditions.

\subsection{Trade-off between Accuracy and Efficiency}\label{exp:tradeoff}
As shown in Figure \ref{fig:tradeoff_accuracy_efficiency}, the trade-off between accuracy and efficiency are investigated under different efficiency budgets among multiple metrics and search algorithms, where the algorithm includes random search and MPEA. As the budget of the efficiency increases, the accuracy of the found subnet tends to improve consistently. It indicates that the trained supernet creates such a subnet space that increasing the budget brings accuracy gain.

In Figure \ref{fig:tradeoff_accuracy_efficiency}, the predicted EER derived from MPEA on the fine-grained space outperforms other methods. However, their actual accuracy has extremely slight improvement and even leads to inferior performance in DCF$_{0.01}$. One reason is that the fine-grained space contains significantly larger subnets (i.e., $\approx 10^{13}$) than the coarse-grained space, which makes it challenging to learn from the number of pairs of accuracy and subnets as equal to the pairs from the coarse-grained. In contrast, the accuracy predictor trained on the coarse-grained space generates approximately equal performance as the actual. It suggests that it is feasible to create an accuracy predictor that learns an appropriate-grained space for finding a subnet with more subtle architectures.

The bottom row in Figure \ref{fig:tradeoff_accuracy_efficiency} shows that the subnets found by random search achieve superior accuracy, e.g., CPU latency ranges 12ms to 16ms, GPU latency ranges 13ms and 17ms, MACs range 0.8G to 1.0G, and parameters range 1.5M to 4.0M. It suggests that under a specific budget of efficiency, the random search can produce efficient architectures while maintaining DCF$_{0.01}$. It implies that an advanced evolutionary algorithm is not necessary to find an efficient network in the proposed supernet.

\section{Sensitivity Analysis}\label{sec:sensitivity}
Several factors are considered to study the impact of the architectural settings on the performance of speaker networks from the proposed supernet, such as training stage, architectural dimensions, and accuracy predictor. Specifically, these subnets in Section \ref{exp:supernet} are evaluated to profile the performance of the supernet at different training stages. Then, a manual grid space is employed to investigate the relationship between accuracy and depth, kernel, width. Finally, the performance of subnets is reported using different accuracy predictors.

\begin{table*}[!t]
  \renewcommand{\arraystretch}{1.00}
  \caption{Accuracy Profile at Different Training Stage on Vox1-O Test}
  \label{tab:performance_stage}
  \centering
  \begin{tabular}{lrrccccc}
  \toprule
  Subnet & MACs & Params & \emph{largest} & \emph{kernel} & \emph{depth} & \emph{width 1} & \emph{width 2} \\
  & & & EER(\%) / DCF$_{0.01}$ & EER(\%) / DCF$_{0.01}$ & EER(\%) / DCF$_{0.01}$ & EER(\%) / DCF$_{0.01}$ & EER(\%) / DCF$_{0.01}$ \\
  \midrule
  $a_\mathrm{max}$ & 1.93G & 7.55M & 1.30 / 0.126 & 1.18 / 0.122 & \textbf{1.11} / 0.120 & 1.29 / 0.131 & 1.44 / 0.163 \\
  $a_\mathrm{Kmin}$ & 1.74G & 6.93M & - & 3.07 / 0.336 & 2.79 / 0.308 & 3.24 / 0.299 & 3.54 / 0.344 \\
  $a_\mathrm{Dmin}$ & 936.82M	& 3.98M & - & - & 3.18 / 0.315 & 3.65 / 0.353 & 3.58 / 0.334 \\
  $a_\mathrm{C1min}$ & 267.44M & 1.25M & - & - & - & 4.12 / 0.405 & 3.98 / 0.360 \\
  $a_\mathrm{C2min}$ & 83.47M & 443.97K	 & - & - & - & - & 5.29 / 0.478 \\
  \bottomrule
  \end{tabular}
\end{table*}

\subsection{Training Stage}
Since the supernet has numerous subnets, it is challenging to determine its performance via a single architecture. To this end, $a_\mathrm{max}$, $a_\mathrm{Kmin}$, $a_\mathrm{Dmin}$, $a_\mathrm{C1min}$, and $a_\mathrm{C2min}$ are summarized to profile the accuracy of the supernet.

Table \ref{tab:performance_stage} reports the accuracy of these subnets with the bounded inference time on the Vox1-O test set. The accuracy varies across different training stages. Specifically, as training tasks undergo from \emph{largest} to \emph{width 2}, the accuracy improves at first but degrades after \emph{width 1}. $a_\mathrm{max}$ derived from the weights of $\emph{depth}$ becomes the optimal solution in EER. It implies that using dynamic kernel and depth during training can benefit the optimization of weights. 
However, the dynamic width training degrades accuracy. It probably results from that pruning the supernet in width removes a large number of parameters that causes the problem of over-regularization \cite{pmlr-v119-xu20e}.

In Table \ref{tab:performance_stage}, the minimal available architecture results in inferior accuracy. For example, the EER of $a_\mathrm{Kmin}$ in \emph{kernel}, $a_\mathrm{Dmin}$ in \emph{depth}, $a_\mathrm{C1min}$ in \emph{width 1}, and $a_\mathrm{C2min}$ in \emph{width 2} generate low accuracy. On the other hand, architecture accuracy can be improved using the progressive training method. Specifically, the accuracy of $a_\mathrm{Kmin}$ in \emph{depth} outperforms that in \emph{kernel}, and the $a_\mathrm{C1min}$ in \emph{width 2} is superior to that in \emph{width 1}. It suggests that for some specific subnets, the progressive training with smaller architectures can achieve accuracy gain. However, $a_{\mathrm{D}_\mathrm{min}}$ in \emph{width 1} has a lower accuracy than that in \emph{depth}. It probably implies a trade-off between pruning smaller subnets and maintaining the accuracy of larger subnets.

\subsection{Architectural Dimensions}
The trained supernet is used to analyze the relationship between accuracy and architectural factors, e.g., network depth, kernel size, and layer width. The grid space is created in the form of $a_\mathrm{grid}(D, K, C) \equiv (D, \{K\}_{i=1}^{D+1}, \{C\}_{i=1}^{D+1} \cup \{3C\})$, where $D$ is chosen from $\{2, 3, 4\}$, $K$ is chosen from $\{1, 3, 5\}$, and $C$ is chosen from $\{128, 136, \dots, 512\}$.

\subsubsection{Network Depth}
\begin{figure}[!t]
  \centering
  \includegraphics[width=8cm]{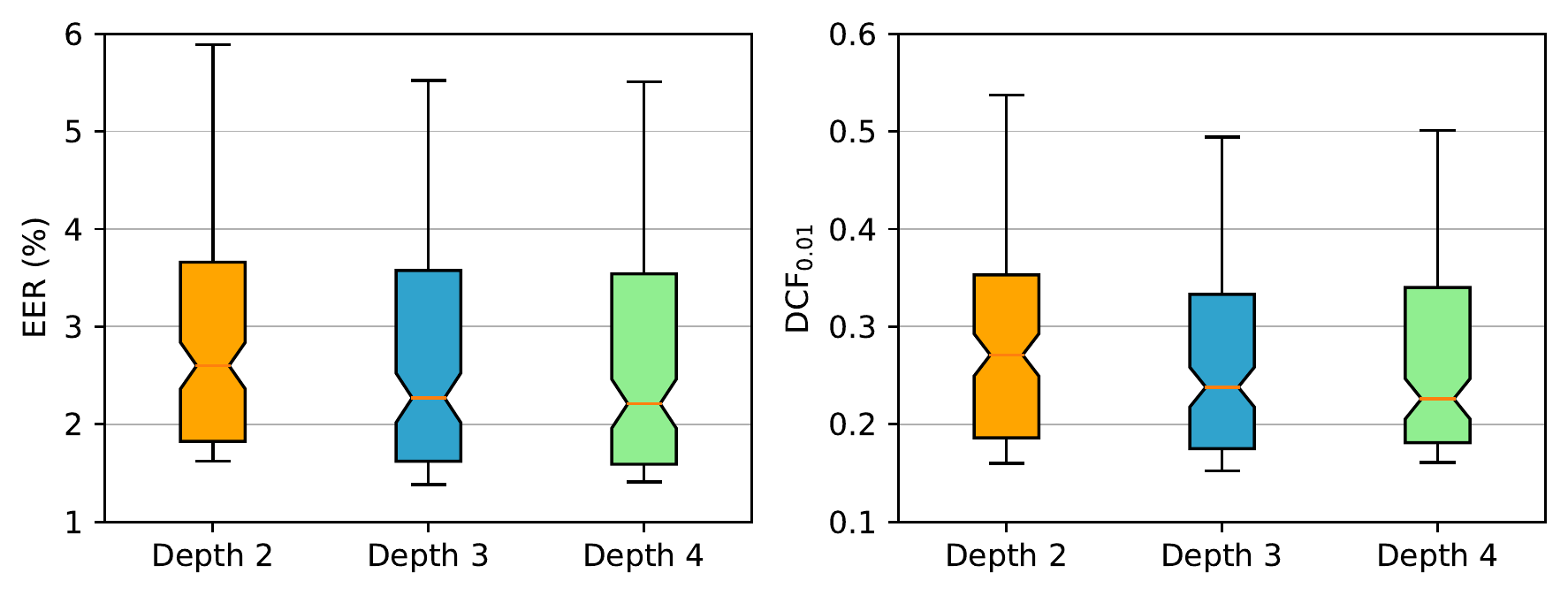}
  \caption{Distribution between accuracy and depth under grid space, where a box denotes the accuracy of architectures of the specific depth.}
  \label{fig:grid_depth}
\end{figure}

In order to investigate the relationship between depth and accuracy, the subnets are sampled from grid space with different depths. As illustrated in Figure \ref{fig:grid_depth}, the networks with the depth of 2 achieve slightly inferior accuracy compared to that with the depth of 3 or 4, and the accuracy of subnets with depths of 3 and 4 are comparable. It suggests that in the trained supernet, different frame-level features contribute to the accuracy of subnets equally approximately.

\subsubsection{Kernel Size}
\begin{figure}[!t]
  \centering
  \includegraphics[width=8cm]{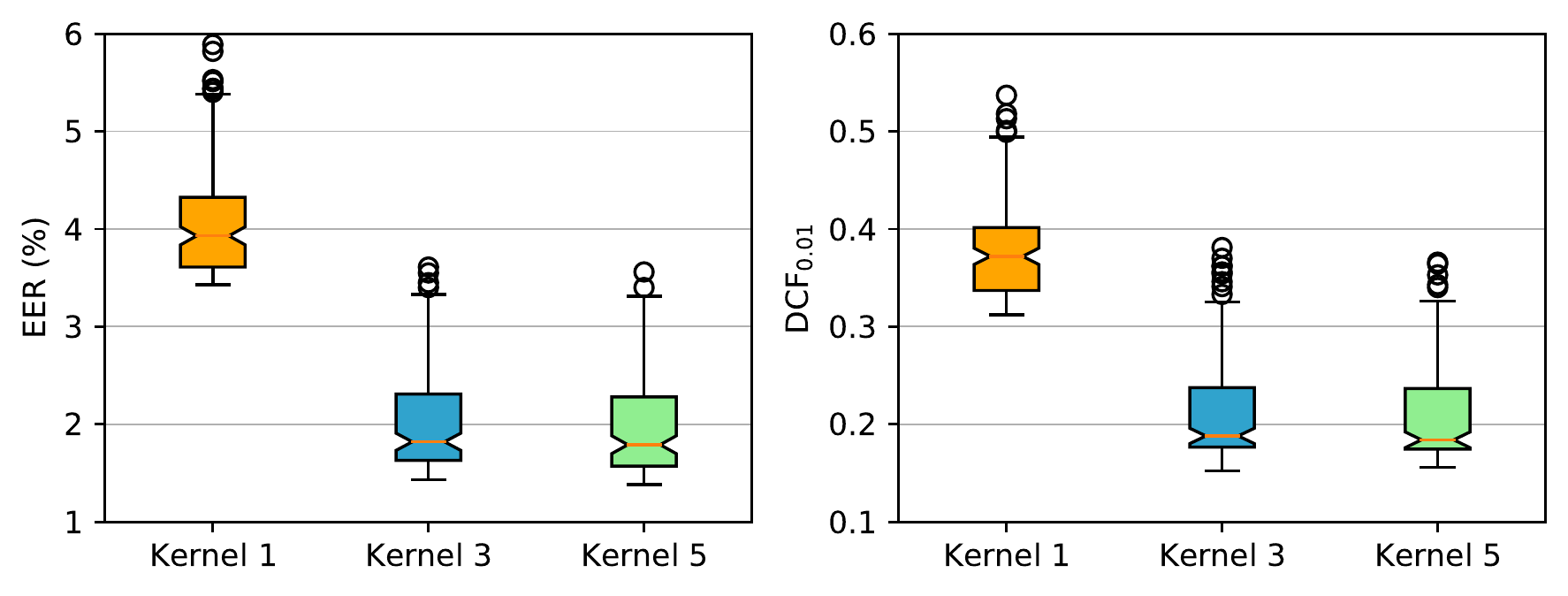}
  \caption{Distribution between accuracy and kernel size under grid space, where a box denotes the accuracy of architectures with a kernel size.}
  \label{fig:grid_kernel}
\end{figure}

As shown in Figure \ref{fig:grid_kernel}, the accuracy of the kernel sizes of 3 and 5 significantly outperforms the kernel size of 1 in EER and DCF$_{0.01}$. It indicates that learning adjacent information helps networks extract discriminative embeddings for speaker recognition. On the other hand, the subnets with a kernel size of 5 have an EER or DCF$_{0.01}$ similar to that with a kernel size of 3, which means that further increasing the receptive field in the dilated form obtains little accuracy gain. Accordingly, reducing kernel size from 5 to 3 leads to smaller architectures with a slight accuracy loss.

\subsubsection{Layer Width}
\begin{figure}[!t]
  \centering
  \includegraphics[width=8cm]{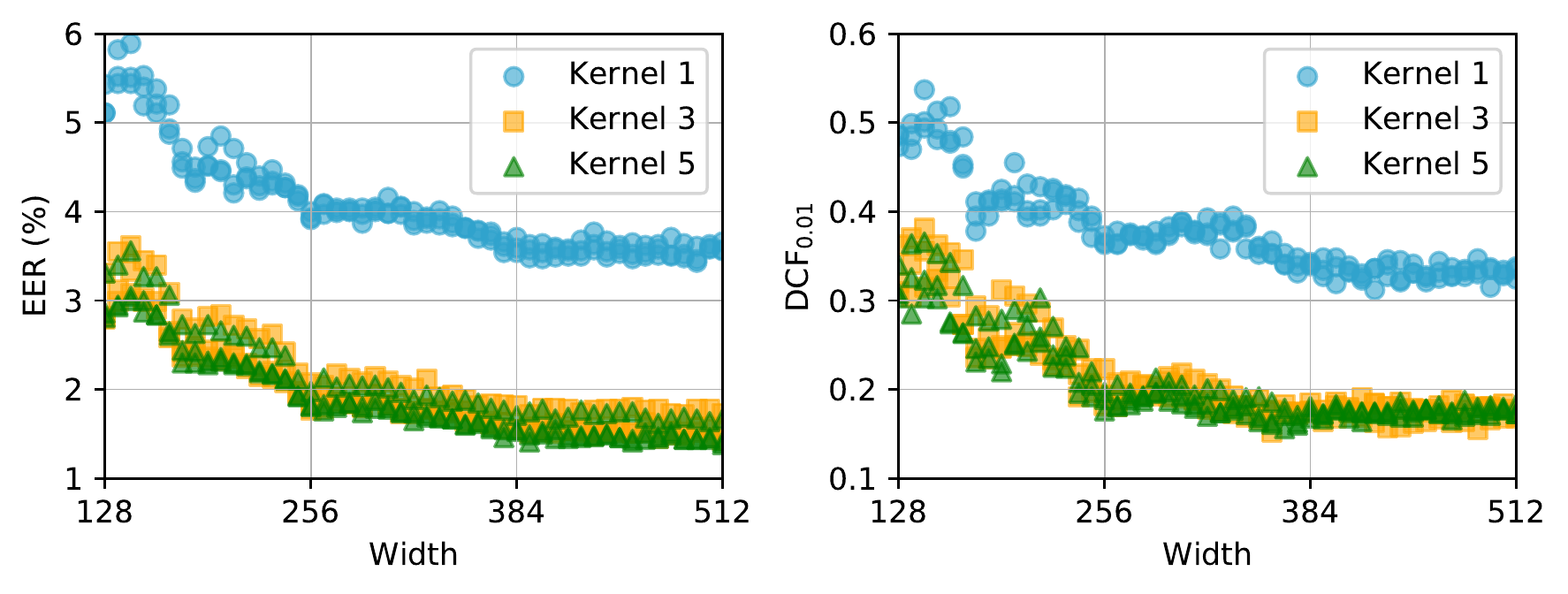}
  \caption{Distribution between accuracy and width under grid space.}
  \label{fig:grid_width}
\end{figure}

Figure \ref{fig:grid_width} shows that the increasing width consistently improves the accuracy of subnets. It indicates a monotonic relationship between accuracy and width. Also, the width of around 384 creates an accuracy approximately equal to the largest one. It suggests that reducing the width from 512 to 384 retains the accuracy of subnets but requires less computation and storage.

\begin{figure}[!t]
  \centering
  \includegraphics[width=8cm]{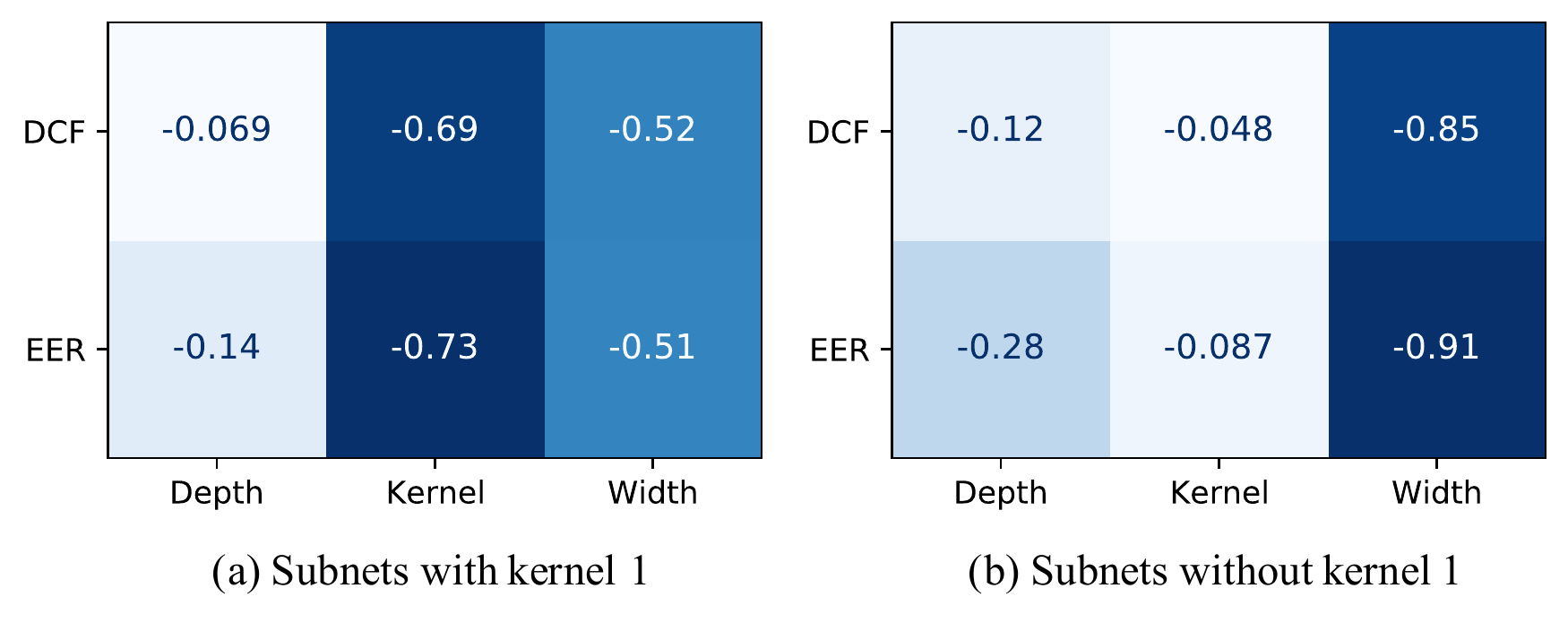}
  \caption{Spearman rank correlation between accuracy and architectural factors, including network depth, kernel size, and layer width.}
  \label{fig:grid_spearman}
\end{figure}

In summary, Table \ref{fig:grid_spearman} shows the Spearman rank correlation between architectural dimensions and the performance of speaker representations, where the higher value denotes the stronger relevance. Figure \ref{fig:grid_spearman}a shows that the network depth has a slight impact on the accuracy of subnets, while the kernel size and the layer width are essential and consistent correlation. However, in the absence of subnets with a kernel size of 1, the kernel sizes of 3 and 5 create similarly low correlation coefficients, as illustrated in Figure \ref{fig:grid_spearman}b, which meets the analysis of kernel sizes. We can conclude that the subnet $a_\mathrm{grid}(3, 3, 384)$ can achieve comparable accuracy to the largest one. As shown in Table \ref{tab:grid_conclusion}, compared with $a_\mathrm{grid}(4, 5, 512)$, the subnet $a_\mathrm{grid}(3, 3, 384)$ reduces 57\% MACs and 55\% parameters and suffers a slight loss in EER.

\subsection{Accuracy Predictor Performance}
\begin{figure}[!t]
  \centering
  \includegraphics[width=8cm]{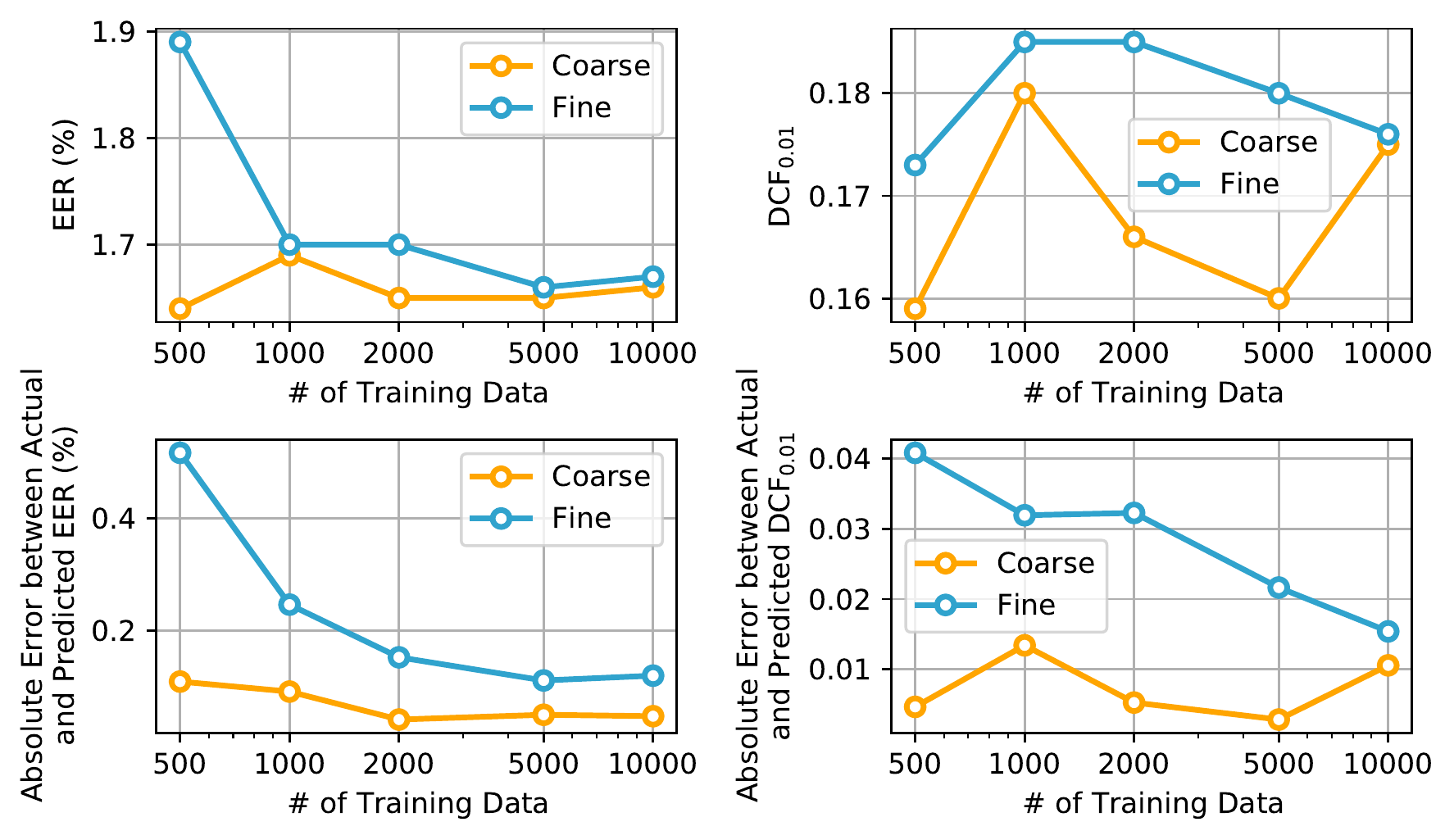}
  \caption{Performance of MPEA-selected subnets with 600M MACs using predictors with different training data.}
  \label{fig:mpea_acc_preformance}
\end{figure}

Figure \ref{fig:tradeoff_accuracy_efficiency} indicates that MPEA achieves different accuracy as the predictor varies.
The accuracy predictor is conducted with different training data to investigate the predicted and actual accuracy relationship. Figure \ref{fig:mpea_acc_preformance} shows the results of the selected subnets with 600M MACs among different accuracy predictors. It illustrates that subnets derived from the coarse-grained space have lower predicted errors and higher recognition accuracy than that from the fine-grained space. It suggests that the coarse-grained space predictor is more robust than the fine-grained space predictor. Considering the sizes of the fine-grained space are $10^{6} \times$ larger than the sizes of the coarse-grained space, it implies that a more subtle space requires an increasing number of training data.

\begin{table}[!t]
  \renewcommand{\arraystretch}{1.00}
  \caption{Performance of Subnets Derived from Grid Search and Analysis on the Vox1-O}
  \label{tab:grid_conclusion}
  \centering
  \begin{tabular}{lrrcc}
  \toprule
  Subnet & MACs & Params & EER(\%) & DCF$_{0.01}$ \\
  \midrule
  $a_\mathrm{grid}(3, 3, 384)$ & 826.11M & 3.42M &  1.54\% & 0.148 \\
  $a_\mathrm{grid}(4, 5, 512)$ & 1.93G & 7.55M & 1.44\% & 0.163 \\
  \bottomrule
  \end{tabular}
\end{table}

\section{Conclusion and Future Work}\label{sec:conclusion}
In this paper, we propose EfficientTDNN to address the problem of searching for the specialized network for speaker recognition that meets  specific constraints. Experimental results on the VoxCeleb dataset show the proposed EfficientTDNN enables a significantly large number of architectural settings ($\approx 10^{13}$) concerning depth, kernel, and width. 
Considering different computation constraints, it achieves 2.20\% EER with 204M MACs, 1.41\% EER with 571M MACs as well as 0.94\% EER with 1.45G MACs, which outperforms D-TDNN, ResNet-20, Dual Attention, H/ASP, ARET-25, ECAPA-TDNN, Fast ResNet-34, and AutoSpeech.
Comprehensive investigations suggest that the trained supernet generalizes cells with more subtle widths and provides a favorable trade-off between accuracy and efficiency by reducing or expanding the size of architectural settings.

In the future, this work could be improved from the following aspects. Firstly, the designed supernet ranging from 83.47M and 1.93G MACs is possibly not suitable for network pruning since the lower bound of efficiency is so low that large subnets have to adapt to smaller ones which may cause over-regularization. Secondly, the sensitivity analysis of architectural factors shows that a kernel size of 1 leads to inferior accuracy but still serves as an option for the architecture search. Thirdly, the fine-grained space provides subtle width options. However, it makes training an accuracy predictor hard. Therefore, follow up work will include the following: (1) Building a supernet that has an appropriately lower bound of network efficiency; (2) Designing an adaptive search method that eliminates options that lead to inferior performance; (3) Establishing a search space that balances the granularities of a sampling space and data requirements for training an accuracy predictor.

\section*{Acknowledgements}
\label{sec:acknowledgements}

The work is partially supported by the National Nature Science Foundation of China (No. 61976160, 61906137, 61976158, 62076184, 62076182, 62072408) and Shanghai Science and Technology Plan Project (No. 21DZ1204800) and Technology research plan project of Ministry of Public and Security (Grant No. 2020JSYJD01). We want to thank our TASLP reviewers for great feedback on the paper.

\ifCLASSOPTIONcaptionsoff
  \newpage
\fi

\bibliographystyle{IEEEtran}
\bibliography{IEEEabrv,ea}

\begin{IEEEbiography}[{\includegraphics[width=1in,height=1.25in,clip,keepaspectratio]{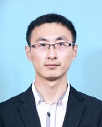}}]{Rui Wang} received B.S. and M.S. degrees both from Zhejiang Sci-Tech University, China, in 2015 and 2018, respectively. He is currently pursuing the Ph.D. degree in Computer
Science and Technology from Tongji University, Shanghai, China, under the supervision of Zhihua Wei. His current research interests focus on Speaker Recognition, Neural Architecture Search, and Self-supervised Learning. He was a Research Intern at Microsoft Research Asia in 2021. He is also a reviewer of IEEE TASLP.
\end{IEEEbiography}

\begin{IEEEbiography}[{\includegraphics[width=1in,height=1.25in,clip,keepaspectratio]{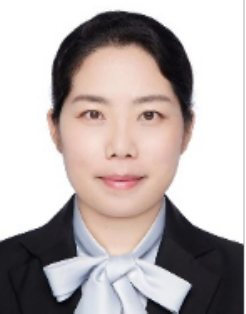}}]{Zhihua Wei} is currently a Professor at Tongji University. She received a Ph.D. degree pattern recognition and intelligent system from Tongji University in China, a Ph.D. degree in Information from Lyon2 University in France, and B.S. and M.S. degrees both in Computer Science from Tongji University in China. Her current research interests include machine learning, natural language processing, and speech processing. She is a member of the granular computing and knowledge discovery Professional Committee of Chinese Artificial Intelligence Society, and a member of the natural understanding professional committee of Chinese Artificial Intelligence Society. 
\end{IEEEbiography}

\begin{IEEEbiography}[{\includegraphics[width=1in,height=1.25in,clip,keepaspectratio]{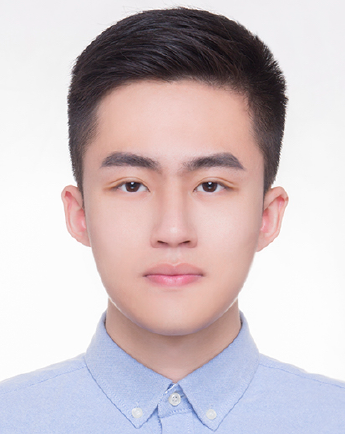}}]{Haoran Duan} received a Distinction M.S. degree in Data Science from Newcastle University, UK, in 2019. After that, he was a research student in OpenLab, Newcastle University, UK, and he was also a research associate at Newcastle University working on deep learning applications. He is currently pursuing a PhD degree in the Department of Computer Science, Durham University. His current research interests focus on the application/theory of deep learning. He is also an active reviewer of CVPR, ECCV, AAAI, BMVC, and Ubicomp.
\end{IEEEbiography}

\vfill
\newpage

\begin{IEEEbiography}[{\includegraphics[width=1in,height=1.25in,clip,keepaspectratio]{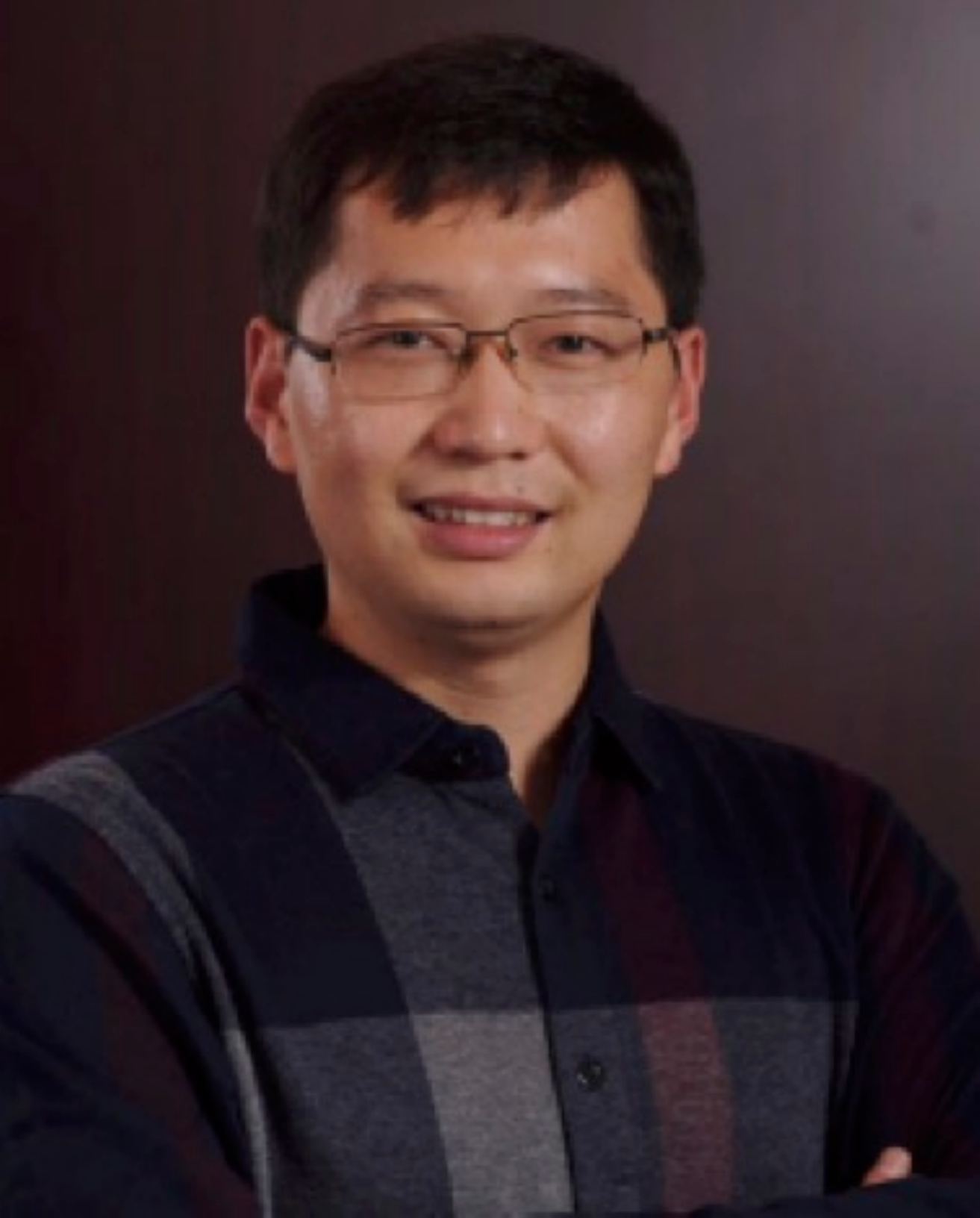}}] {Shouling Ji} is a ZJU 100-Young Professor in the College of Computer Science and Technology at Zhejiang University and a Research Faculty in the School of Electrical and Computer Engineering at Georgia Institute of Technology (Georgia Tech). He received a Ph.D. degree in Electrical and Computer Engineering from Georgia Institute of Technology, a Ph.D. degree in Computer Science from Georgia State University, and B.S. (with Honors) and M.S. degrees both in Computer Science from Heilongjiang University. His current research interests include Data-driven Security and Privacy, AI Security and Big Data Analytics. He is a member of ACM, IEEE, and CCF and was the Membership Chair of the IEEE Student Branch at Georgia State University (2012-2013). He was a Research Intern at the IBM T. J. Watson Research Center. Shouling is the recipient of the 2012 Chinese Government Award for Outstanding Self-Financed Students Abroad and eight best paper awards.
\vspace{-8pt}\end{IEEEbiography}

\begin{IEEEbiography}[{\includegraphics[width=1in,height=1.25in,clip,keepaspectratio]{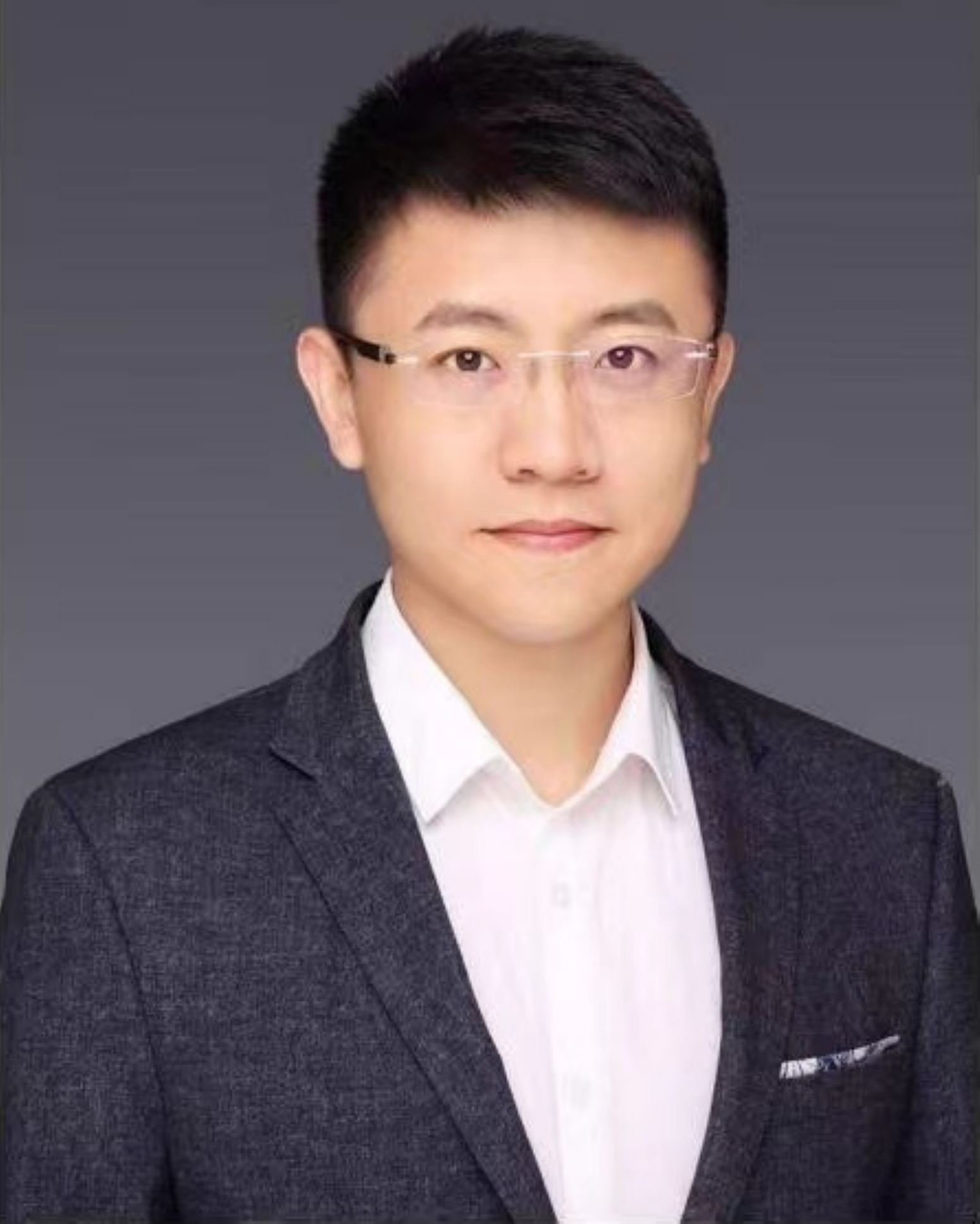}}] {Yang Long} is an Assistant Professor in the Department of Computer Science, Durham University. He is also an MRC Innovation Fellow aiming to design scalable AI solutions for large-scale healthcare applications. His research background is in the highly interdisciplinary field of Computer Vision and Machine Learning. While he is passionate about unveiling the black-box of AI brain and transferring the knowledge to seek Scalable, Interactable, Interpretable, and sustainable solutions for other disciplinary researches, e.g. physical activity, mental health, design, education, security, and geoengineering. He has authored/coauthored 30+ top-tier papers in refereed journals/conferences such as IEEE TPAMI, TIP, CVPR, AAAI, and ACM MM, and holds a patent and a Chinese National Grant.
\vspace{-8pt}\end{IEEEbiography}

\begin{IEEEbiography}[{\includegraphics[width=1in,height=1.25in,clip,keepaspectratio]{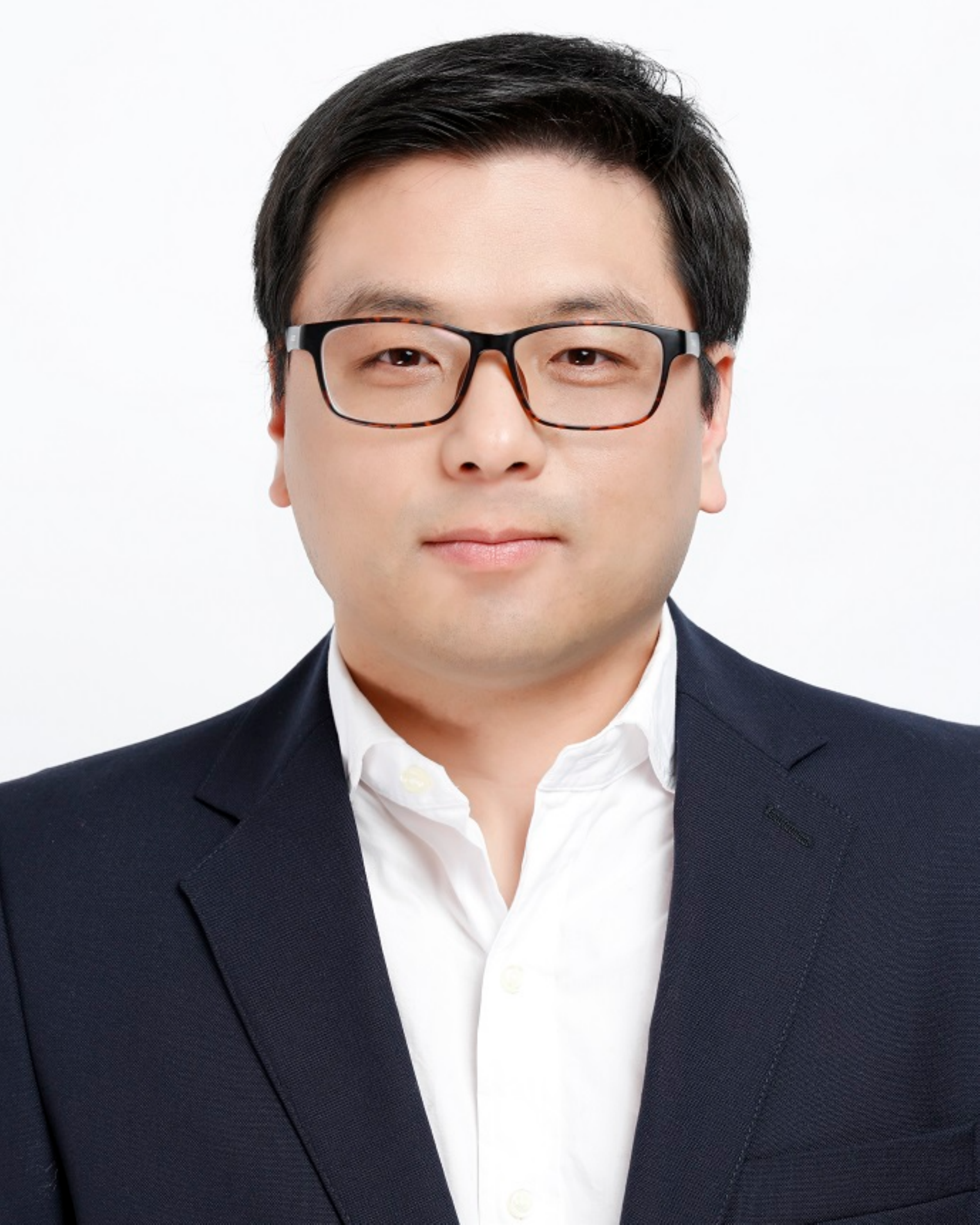}}] {Zhen Hong} received a B.S. degree of Computer Science \& Technology and Computing from Zhejiang University of Technology (China) and University of Tasmania (Australia) in 2006, respectively, and a Ph.D. from the Zhejiang University of Technology in 2012. He was an associate professor with the Faculty of Mechanical Engineering \& Automation, Zhejiang Sci-Tech University, China. He is now a professor and vice-dean with the Institute of Cyberspace Security, and College of Information Engineering, Zhejiang University of Technology, China. Dr. Hong has visited at the Sensorweb Lab, Department of Computer Science, Georgia State University in 2011. He also has been at CAP Research Group, School of Electrical \& Computer Engineering, Georgia Institute of Technology as a research scholar in 2016 to 2018. His research interests include cyber-physical systems, Internet of things, wireless sensor networks, cybersecurity, AI-based methods, and data analytics. He received the first Zhejiang Provincial Young Scientists Title in 2013 and the Zhejiang Provincial New Century 151 Talent Project (The Third-Level) in 2014. He is a member of IEEE/ACM, CCF and senior member of CAA, and serves on the Youth Committee of Chinese Association of Automation and Blockchain Committee and AC of CCF YOCSEF, respectively.
\end{IEEEbiography}
\vfill
\end{document}